\documentclass[aps,reprint,twocolumns,pre,superscriptaddress,floatfix,notitlepage,nofootinbib]{revtex4-1}
\usepackage{amssymb,amsmath,amsfonts} 
\usepackage{mathtools}
\usepackage{physics}
\usepackage{array}
\usepackage{graphicx,epsfig,xcolor}
\usepackage[colorlinks=true,citecolor=blue,linkcolor=red]{hyperref}
\usepackage{newtxtext}

\newcommand{\beq}{\begin{equation}}
\newcommand{\eeq}{\end{equation}}
\begin{document}

\newcommand{\hatmath}[1]{\hat{\mathcal{#1}}} 
\newcommand{\ddt}[1]{\frac{\textrm{d}{#1}}{\textrm{d}t}}
\newcommand{\tx}[1]{\textrm{#1}} 

\title{Constants of motion characterizing continuous symmetry-broken phases}

\author{\'{A}ngel L. Corps}
    \email[]{corps.angel.l@gmail.com}
    \affiliation{Instituto de Estructura de la Materia, IEM-CSIC, Serrano 123, E-28006 Madrid, Spain}
    \affiliation{Grupo Interdisciplinar de Sistemas Complejos (GISC),
Universidad Complutense de Madrid, Av. Complutense s/n, E-28040 Madrid, Spain}

\author{Jorge Dukelsky}
    \email[]{dukelsky@gmail.com}
    \affiliation{Instituto de Estructura de la Materia, IEM-CSIC, Serrano 123, E-28006 Madrid, Spain}
    
\author{Armando Rela\~{n}o}
    \email[]{armando.relano@fis.ucm.es}
    \affiliation{Grupo Interdisciplinar de Sistemas Complejos (GISC),
Universidad Complutense de Madrid, Av. Complutense s/n, E-28040 Madrid, Spain}
    \affiliation{Departamento de Estructura de la Materia, F\'{i}sica T\'{e}rmica y Electr\'{o}nica, Universidad Complutense de Madrid, Av. Complutense s/n, E-28040 Madrid, Spain}
\begin{abstract}
We present a theory characterizing the phases emerging as a consequence of continuous symmetry-breaking in quantum and classical systems. In symmetry-breaking phases, dynamics is restricted due to the existence of a set of conserved charges derived from the order parameter of the phase transition. Their expectation values are determined by the privileged direction appearing in the ordered phase as a consequence of symmetry breaking, and thus they can be used to determine whether this direction is well defined or it has quantum fluctuations. Our theory is numerically exemplified via the two-dimensional limit of the vibron model, a fully connected system invariant under a rotation operator which generates the continuous symmetry-breaking. 

\end{abstract}

\date{\today} 
\maketitle

\section{Introduction}

Spontaneous symmetry-breaking (SSB) is a key concept in theoretical physics \cite{Arodz2012,Beekman2019}. It plays a prominent role in phase transitions in condensed matter physics \cite{Landau1937,Landau1950}, the generation of mass via the Higgs mechanism \cite{Englert1964,Higgs1964,Guralnik1964}, or the early stages in the evolution of the Universe \cite{Albrecht1982,Kibble1976}. Continuous SSB appears in nature when the physical laws are invariant under a continuous transformation, but equilibrium states, typically below a certain critical temperature, are not. They are linked to a number of physical phenomena, like the existence of Nambu-Goldstone modes \cite{Nambu1960,Goldstone1961}, the presence of off-diagonal long-range order \cite{Yang1962}, and the appearance of a set of energy levels which collapse onto the ground state in the thermodynamic limit, the so-called Anderson tower of states \cite{Anderson1952,Bernu1992,Azaria1993,Bernu1994,Koma1994,Tasaki2019}.

Besides these features, the best analogy to understand continuous symmetry breaking is the paradigmatic classical example of a pencil situated vertically and upside down over a table. Even though this initial condition and the physical laws governing its evolution are symmetric under any rotation, a tiny perturbation is enough to make the pencil fall down and point towards a particular direction, therefore breaking the rotational symmetry. This entails two observable consequences. First, the length of the projection of the pencil over the table changes from zero in the symmetric initial condition, to a non-zero value in the final, symmetry-broken, state. And second, we find the emergence of a privileged direction, identified, for example, by the angle sustained by the pencil tip with respect to a given (arbitrary) direction.  

This analogy suggests that continuous symmetry breaking can be identified by an order parameter in classical and quantum systems: a physical magnitude which is equal to zero in symmetric equilibrium states and different from zero in symmetry-breaking ones, and that is somehow linked to the privileged direction appearing as a consequence of the symmetry breaking \cite{Landau1937}. In the famous Bose-Hubbard model \cite{Gersch1963}, this role is played by the on-site Bose-field operator, $\hat{b}_k$ \cite{Sachdev}. The model Hamiltonian is invariant under a global U(1) rotation, $\hat{b}_k \to \hat{b}_k \, \textrm{e}^{i \phi}$, and therefore $\langle \hat{b}_k \rangle \neq 0$ if and only if this symmetry is broken. In such a case, $\langle \hat{b}_k \rangle$ is a complex-valued magnitude, whose phase, $\phi$, determines the 'direction' in which the symmetry is broken, and whose modulus quantifies the degree of symmetry breaking. Unfortunately, $\hat{b}_k$ is not an hermitian operator, and therefore it cannot be directly measured in experiments. As a consequence, the ordered phase is usually identified instead by means of other magnitudes indirectly related to the symmetry-breaking, like the ground state gap or the presence of long-range order in the lattice \cite{Fisher1989,Greiner2002}. In the antiferromagnetic Heisenberg model \cite{Anderson1952}, the usual order parameter is the staggered magnetization in any spatial direction, $\hat{M}_{\vec{u}} = \sum_i (-1)^i \vec{u} \cdot \vec{\sigma}_i$ \cite{Dyson1978}, which can be measured in experiments \cite{Mazurenko2017}. However, this apparently simpler situation suffers from its own difficulties. In each experimental realization, SSB can happen in any direction, and thus it is not possible to select in advance the right direction $\vec{u}$ for the order parameter. And even though SSB can be inferred from the probability distribution of the staggered magnetization on any direction \cite{Mazurenko2017}, there is not a clear link between this result and the actual privileged direction appearing as a consequence of SSB. Furthermore, none of these choices for the order parameter provides a satisfactory answer to the following question: is the privileged direction a {\em classical} magnitude, well defined in each realization? Or is it a {\em quantum} magnitude, with quantum fluctuations, as expected in a quantum phase transition \cite{Sachdev,Larson2017}?

In this paper, we focus on the dynamical consequences of entering a continuous symmetry-breaking phase. Generalizing previous results about discrete symmetry-breaking phases \cite{Corps2021,Corps2022,Corps2023,Corps2023arxiv}, we propose that a continuous symmetry-breaking phase is characterized by the appearance of a number of Hermitian operators which are constants of motion only within the symmetry-breaking phase. We derive a general theory showing that their expectation values depend on the privileged direction emerging as a consequence of the SSB and on its quantum fluctuations. As a test-bed for our general theory, we perform numerical simulations on the two-dimensional limit of the vibron model \cite{Khalouf2021,PerezBernal2005,Iachello2003,Iachello1996,PerezBernal2008,Khalouf2022,PerezBernal2010,Novotny2023}. This is an algebraic, fully connected model with important applications in the description of the behavior of certain molecules. Structurally, it is two-level interacting model with three bosonic species. The model has U(1) invariance driven by a continuous rotational symmetry, and thus it is presented as a powerful system to test our theoretical results.

This paper is organized as follows. The core of our theory is laid out in Sec. \ref{sec:theory}; in particular, the general setup that we consider in this work is presented in Sec. \ref{sec:setup}, the identification of constants of motion from order parameters is explained in Sec. \ref{sec:constants}, and Sec. \ref{sec:insight} is devoted to the physical role played by the constants of motion previously defined. Details our model Hamiltonian are found in Sec. \ref{sec:modelham}. Its classical limit is then discussed in Sec. \ref{sec:classicallimit}, and the classical interpretation of symmetry-breaking is presented in Sec. \ref{sec:classicaldynamics}. Sec. \ref{sec:quantum} contains our numerical results on the quantum model, with Sec. \ref{sec:spectralproperties} delving into the role played by its spectral properties in the emergence of symmetry-breaking, Sec. \ref{sec:quantumquenches} focusing on non-equilibrium dynamics generated by a quantum quench, and Sec. \ref{sec:information} presenting a statistical model connecting the expectation values of the quantum constants of motion with a symmetry-breaking angle with classical significance. Finally, our conclusions are gathered in Sec. \ref{sec:conclusions}.

\section{Theory}\label{sec:theory}

In this section, we develop a framework for equilibrium states within a continuously symmetry broken phase. As we will see later, the main trademark of such equilibrium states is their dependence on a set of charges that do no commute with the operator that generates the continuous symmetry. Our focus is not on a particular physical model but on a wide class of models with common properties. 

\subsection{Generic setup for continuous symmetry-breaking equilibrium states}\label{sec:setup}

Let us consider a time-independent Hamiltonian, $\hat{H}$, and a unitary operator,
\beq\label{eq:Rsym}
\hat{R}(\alpha)=e^{i\alpha \hat{L}},\,\,\,\alpha\in[0,2\pi),
\eeq
which may be understood as a rotation generated by an Hermitian operator,  $\hat{L}^{\dagger}=\hat{L}$. We further assume that both $\hat{H}$ and $\hat{L}$ have a discrete spectrum, and that $\hat{H}$ is invariant under $\hat{R}(\alpha)$,
\beq
\hat{R}(\alpha)\hat{H}\hat{R}^{\dagger}(\alpha)=\hat{H},\,\,\forall\alpha\in[0,2\pi).
\eeq
This implies that $\hat{R}(\alpha)$ is a \textit{continuous symmetry}. The commutation $[\hat{R}(\alpha),\hat{H}]=0$, $\forall \alpha$, implies that $[\hat{L},\hat{H}]=0$, i.e., $\hat{L}$ is a conserved quantity. Without loss of generality, we consider that the eigenvalues of $\hat{L}$ are $\ell\in\mathbb{Z}$. Since $\hat{H}$ and $\hat{L}$ are commuting, there exists a common eigenbasis for these operators. In particular, the Hamiltonian eigenbasis, $\{\ket{E_{n}^{\ell}}\}$, diagonalizes $\hat{L}$,
\begin{subequations}
\label{eq:eigenvalues}
\begin{align}
\hat{H}\ket{E_{n}^{\ell}}=E_{n}^{\ell}\ket{E_{n}^{\ell}},\\\hat{L}\ket{E_{n}^{\ell}}=\ell\ket{E_{n}^{\ell}},
\end{align}
\end{subequations}
where $n=0,1,\ldots$ and $\ell\in\mathbb{Z}$. In a basis where $\ell$ is a good quantum number, the full Hilbert space can be decomposed into the sum of all $\ell$-sectors, $\mathcal{H}=\bigoplus_{\ell\in\mathbb{Z}}\mathcal{H}_{\ell}$, i.e., $\hat{H}$ is block diagonal with blocks of dimension $D_{\ell}$. 

An equilibrium state is represented by a density matrix, $\hat{\rho}$, that remains constant over time, $\hat{\rho}(t)=\hat{\rho}(0)\equiv \hat{\rho}$, i.e.,
\beq
\hat{U}(t)\hat{\rho}\hat{U}^{\dagger}(t)=\hat{\rho},\,\,\,
\eeq
where $\hat{U}(t)=e^{-i\hat{H}t}$ ($\hbar=1$) is the time evolution operator. This implies that a statistical ensemble accounting for all the possible equilibrium states may depend on all the relevant constants of motion, since the unitary evolution preserves the initial expectation values of all of them \cite{Jaynes1957a,Jaynes1957b,Rigol2007}. In our case, it must be a function of the Hamiltonian itself, $\hat{H}$, and the generator of the continuous symmetry, $\hat{L}$, but it may also depend on a larger set of operators. 

To identify the set of operators that we need to take into account, we start by distinguishing two types of equilibrium states depending on their behavior under the symmetry $\hat{R}(\alpha)$. On the one hand, \textit{symmetric} equilibrium states obey the Hamiltonian symmetry,
\beq
\hat{R}(\alpha)\hat{\rho}_{\textrm{S}}\hat{R}^{\dagger}(\alpha)=\hat{\rho}_{\textrm{S}},\,\,\forall \alpha\in[0,2\pi).
\eeq
On the other hand, we have \textit{symmetry-breaking} equilibrium states, which are not invariant under $\hat{R}(\alpha)$ at least for a given $\alpha$,
\beq
\label{eq:rhoSB}
\exists\alpha\in[0,2\pi)\,\,:\,\,\hat{R}(\alpha)\hat{\rho}_{\textrm{SB}}\hat{R}^{\dagger}(\alpha)\neq \hat{\rho}_{\textrm{SB}}.
\eeq

We begin with this last kind of equilibrium states. A necessary and sufficient condition for them is the presence of non-zero matrix elements which are non-diagonal in the eigenbasis common to the Hamiltonian, $\hat{H}$, and the operator generating the continuous symmetry, $\hat{L}$. 
This implies that:

\textit{There exist symmetry-breaking equilibrium states if, and only if, there exist constants of motion which are not commuting with $\hat{L}$}. 

Therefore, any statistical ensemble describing symmetry-breaking equilibrium states must depend on a number of such constants of motion,
\beq
\hat{\rho}_{\textrm{SB}} = \hat{\rho}_{\textrm{SB}} (\hat{H}, \hat{L}, \hat{C}_1, \hat{C}_2 \ldots),
\eeq
where $\hat{C}_1$, $\hat{C}_2,\ldots$ are constants of motion such that $[\hat{C}_{n},\hat{L}]\neq 0$. Our main aim is to build a set of constants of motion that play this role. 

We note that these facts require the existence of degenerate energy levels. In this work, and without loss of generality, we will assume that all degenerate energy levels belonging to the same eigenspace have the same quantum number $n$ in Eqs. \eqref{eq:eigenvalues}, and that they are distinguished by different values of the $\ell$ quantum number.

This situation typically arises when a continuous symmetry-breaking phase transition takes place. It is worth noting that a real phase transition only happens in thermodynamic limit (TL); in finite-size systems, there are usually no degeneracies in the spectrum. Notwithstanding, this is not incompatible with the observation of symmetry-breaking states in such systems, at least during quite long periods of time. The usual explanation applied to continuous symmetry-breaking quantum phase transitions, is the existence of a \textit{thin spectrum} or a \textit{tower of states}: a set of very close, symmetric energy levels collapsing to the ground state in the TL; linear combinations of these generate symmetry-breaking states \cite{Anderson1952,Bernu1992,Azaria1993,Bernu1994,Koma1994,Tasaki2019}. We will see later on that the same phenomenon explains why the constants of motion noncommuting with $\hat{L}$ are only approximate in finite-size systems, and become exact in the TL; as a consequence, symmetry-breaking equilibrium states, $\hat{\rho}_{\textrm{SB}}$, are expected to have a finite lifetime in finite-size systems, which increases with system size.

By contrast, symmetric equilibrium states can only depend on constants of motion commuting with $\hat{L}$. Hence, in this work we will assume that $\hat{H}$ and $\hat{L}$ are the only relevant constants of motion for symmetric equilibrium states,
\beq
\hat{\rho}_{\textrm{S}} = \hat{\rho}_{\textrm{S}} (\hat{H}, \hat{L}).
\eeq

Under these circumstances, we observe that the \textit{dynamics changes qualitatively upon crossing the critical point of a continuous symmetry-breaking phase transition}. In the disordered phase, all the equilibrium states are symmetric; therefore, the only relevant constants of motion are the Hamiltonian itself and the operator generating the continuous symmetry. The main trademark of the ordered phase, however, is the existence of symmetry-breaking equilibrium states. This implies that a new set of constants of motion appear when the system enters this phase, and therefore more information about the initial state is preserved by the unitary evolution.

\subsection{Constants of motion from order parameters}\label{sec:constants}

Quantum phases are normally identified by means of order parameters. The aim of this section is to take advantage of them to define a set of operators which are constant only in the symmetry-breaking phase.

We focus on the simplest version of continuous symmetry breaking, in which the privileged direction emerging in the ordered phase can be identified by means of a single angle. Hence, we define an \textit{order parameter} as an operator, $\hat{O}$, such that
\begin{subequations}
\begin{align}
&\Tr[\hat{\rho}_{\textrm{SB}}\hat{O}]\neq0,\\
\label{eq:rotation}
&\hat{R}(\alpha)\hat{O}\hat{R}^{\dagger}(\alpha)=e^{i\alpha}\hat{O},\,\,\alpha\in[0,2\pi).
\end{align}
\end{subequations}
As a paradigmatic example, we refer to the on-site Bose-field operator for the Bose-Hubbard model \cite{Sachdev}. As shown in Eq. \eqref{eq:rotation}, a rotation introduces a global phase in $\hat{O}$, which can be used to identify the symmetry-breaking direction. Observe that this implies that $\hat{O}$ is non Hermitian, so it does not represent a physically measurable quantity. It is worth to note that there exist more complicated forms of continuous symmetry breaking, like in the antiferromagnetic Heisenberg model, where the privileged direction is determined by two angles, and thus a vectorial order parameter is necessary to play the same role. 

The previous facts imply that there also exist an infinite family of operators, $\{\hat{O}^{n}\}_{n\in\mathbb{N}}$, with analogous properties,
\begin{align}
\hat{R}(\alpha)\hat{O}^{n}\hat{R}^{\dagger}(\alpha)=e^{in\alpha}\hat{O}^{n},\,\,n\in\mathbb{N},\,\,\alpha\in[0,2\pi),
\end{align}
and similarly for $\hat{O}^{\dagger}$. 

All of them can play the role of the order parameter. Thus, they are the starting point of our theory, which rests on the hypotheses that we formulate below.

\textbf{(H1).--} There exists a Hamiltonian subspace $\mathcal{H}_{SB}\subseteq\mathcal{H}$, $\mathcal{H}_{SB}\neq\{\emptyset\}$, spanned by the eigenstates of $\hat{H}$ and $\hat{L}$, which contains symmetry-broken equilibrium states, i.e., there exist $\hat{\rho}_{\textrm{SB}}$ such that $\hat{R}(\alpha)\hat{\rho}_{\textrm{SB}}\hat{R}^{\dagger}(\alpha)\neq\hat{\rho}_{\textrm{SB}}$ $\forall \alpha$. Note, however, that there may exist symmetric states in $\mathcal{H}_{SB}$. Typically, $\mathcal{H}_{SB}$ is spanned by the eigenstates of $\hat{H}$ with eigenenergies below a critical energy, related to a critical temperature. We will assume that the rest of the Hilbert space is symmetric, so $\mathcal{H} = \mathcal{H}_S \oplus \mathcal{H}_{SB}$, and that every equilibrium state within $\mathcal{H}_S$ is symmetric.

It is worth to note that this hypothesis is much more restrictive than Eq. \eqref{eq:rhoSB}, which only requires that $\hat{\rho}_{\textrm{SB}}$ does not remain invariant under a particular rotation to be considered as symmetry breaking. Hence, the theory we develop in this section is only applicable to this restrictive situation. In Sec. \ref{sec:insight} we comment on the existence of equilibrium states in which the continuous symmetry is only partially broken, and how the theory is modified under these circumstances.

The next hypothesis deals with the conservation laws associated to the symmetry-breaking subspace.

\textbf{(H2).--} Consider an initial symmetry-breaking initial state, $\ket{\psi(t=0)}\in\mathcal{H}_{SB}$. Then, the symmetry remains broken throughout the whole time evolution,
\beq\label{eq:H2}
\langle \hat{O}(t)\rangle\equiv \bra{\psi(t)}\hat{O}\ket{\psi(t)}=r (t)e^{i\alpha}\in\mathbb{C},
\eeq
where $\ket{\psi(t)}=\hat{U}(t)\ket{\psi(0)}$, and $r(t)>0$, $\forall t$. The angle 
$\alpha$ remains constant as its value becomes fixed due to symmetry-breaking, $\alpha(t)=\alpha(0)\equiv \alpha$, $\forall t$. The condition $r(t)>0$ is necessary for $\hat{O}$ to be a valid order parameter: if there exists $t_*$ such that $r(t_*)=0$, then $\langle \hat{O} (t_*) \rangle =0$, and therefore $\hat{O}$ does not distinguish symmetric and symmetry-breaking states at this particular time. Furthermore,  \eqref{eq:H2} implies that
$
\Tr[\hat{\rho}_{\textrm{SB}}\hat{O} ]\neq0.
$

This hypothesis is easy to interpret in the \textit{classical} realm. Returning to the paradigmatic example of a pencil situated vertically and upside down, and subjected to interactions which are invariant under any rotation, hypothesis (H2) states that the symmetry-breaking direction where the pencil falls down is unaltered by the time evolution. Furthermore, Eq. \eqref{eq:H2} establishes a way to determine this direction, since, in the classical limit, the expectation value of $\hat{O}$ is the only possible experimental result of measuring the corresponding observable. Therefore, it is reasonable to conclude that the angle $\alpha$ acts as a classical constant of motion.

On the other hand, the \textit{quantum} interpretation of this hypothesis is more involved. It is logical to assume that the constancy of the angle $\alpha$ must imply the constancy of, at least, a particular observable. However, it is not easy to derive it from the order parameter. First of all, $\hat{O}$ is a non-Hermitian operator, and therefore it is not an observable. As a consequence, we cannot interpret the expectation value in Eq. \eqref{eq:H2} as an average over a set of possible experimental outcomes, each one corresponding to a different angle, $\langle \hat{O} \rangle \neq \sum_n p_n O_n$, $p_{n}$ denoting the probability of measuring the eigenvalue $O_{n}$ of $\hat{O}$. A possible way to overcome this difficulty is to define two Hermitian operators from $\hat{O}$ and $\hat{O}^{\dagger}$,  
\begin{subequations}
\begin{align}
\hat{X}_{1}\equiv \frac{\hat{O}+\hat{O}^{\dagger}}{2},\\
\hat{Y}_{1}\equiv \frac{\hat{O}-\hat{O}^{\dagger}}{2i},\
\end{align}
\end{subequations}
whose instantaneous expectation values are
\begin{subequations}
\begin{align}
\langle \hat{X}_{1} (t)\rangle=r(t)\cos(\alpha),\\
\langle \hat{Y}_{1} (t)\rangle=r(t)\sin(\alpha).
\end{align}
\end{subequations}
Therefore, the angle $\alpha$ can be inferred, up to a global $\pi$ rotation, from
\begin{equation}
\label{eq:tanalpha}
\frac{\langle \hat{Y}_1(t) \rangle}{\langle \hat{X}_1(t) \rangle} = \tan \alpha.
\end{equation}
Hence, a promising solution to our problem is to define a new observable
\begin{equation}
\hat{Z}_1 = \frac{(\hat{X}_1)^{-1} \hat{Y}_1 + \hat{Y}_1 (\hat{X}_1)^{-1}}{2},
\end{equation}
where the summation is done to take into account the possibility that  $[\hat{X}_1, \hat{Y}_1] \neq 0$. Unfortunately, this proposal has two important issues:

(i) Eq. \eqref{eq:tanalpha} does not guarantee that
\begin{equation}
\langle \hat{Z}_1 \rangle = \tan \alpha.
\end{equation}
Therefore, without (at least) one further hypothesis, we cannot state that $\hat{Z}_1$ is a constant of motion within $\mathcal{H}_{SB}$, from which we can deduce the probability of observing a given symmetry-breaking angle $\alpha$ in a measurement.

(ii) More importantly, the existence of $(\hat{X}_1)^{-1}$ is not guaranteed either. For example, this proposal does not always work for the Bose-Hubbard model: the bosonic operator $( \hat{b}_k + \hat{b}_k^{\dagger})^{-1}$ only exists if the number of particles is odd.

In this paper, we follow a different path. Our aim is to generalize former results for discrete $\mathbb{Z}_2$ symmetries, establishing that we need two extra constants of motion, each one having just two possible eigenvalues, $+1$ and $-1$, to describe all the possible symmetry-breaking equilibrium states \cite{Corps2021,Corps2023arxiv}. To do so, we first build a bridge to join the classical and the quantum descriptions. We define a fictitious\footnote{This kind of states cannot exist in real quantum systems. The closer ones are coherent states. As they are eigenstates of the annihilation operator, we can consider that they fulfill \eqref{eq:clasico1}, but not \eqref{eq:clasico2}. However, they can be asymptotically approached in fully-connected systems, for which we can ignore the quantum fluctuations in the thermodynamic limit.} classical state, $\ket{\alpha}$ as an eigenstate of both $\hat{O}$ and $\hat{O}^{\dagger}$,
\begin{eqnarray}
\label{eq:clasico1}
\hat{O} \ket{\alpha} &=& r \textrm{e}^{i \alpha} \ket{\alpha}, \\
\label{eq:clasico2}
\hat{O}^{\dagger} \ket{\alpha} &=& r \textrm{e}^{- i \alpha} \ket{\alpha},
\end{eqnarray}
where $r$ does not depend on the angle $\alpha$, since we can go from a classical state $\ket{\alpha}$ to another by means of a rotation $\hat{R}(\alpha')$. The trademark of these states is the total absence of quantum fluctuations. 

Then, we formulate two consequences of hypothesis (H1) and (H2), which hold in these states:

{\bf (C1).--} Hypothesis (H2) implies that
\beq\label{eq:H2b}
\langle \hat{O}^n (t)\rangle\equiv \bra{\alpha(t)}\hat{O}^n\ket{\alpha(t)}= [ r (t)]^n e^{i n \alpha}\in\mathbb{C}, \; \forall n \in \mathbb{N},
\eeq
and, therefore,
\beq
\Tr[\hat{\rho}_{\textrm{SB}}\hat{O}^n ]\neq0,
\eeq
if $r_{\alpha}(t)>0$. 

From (H1), (H2) and (C1), we can derive two infinite families of Hermitian operators, $\{\hat{X}_{n}\}_{n\in\mathbb{N}}$ and $\{\hat{Y}_{n}\}_{n\in\mathbb{N}}$, where
\begin{subequations}\label{eq:XnYn}
\begin{align}
\hat{X}_{n}\equiv \frac{\hat{O}^{n}+(\hat{O}^{\dagger})^{n}}{2},\\
\hat{Y}_{n}\equiv \frac{\hat{O}^{n}-(\hat{O}^{\dagger})^{n}}{2i}.\
\end{align}
\end{subequations}
The instantaneous expectation values of these observables are given by 
\begin{subequations}
\label{eq:expectationXY}
\begin{align}
\langle \hat{X}_{n}(t)\rangle=[r (t)]^n \cos(n \alpha),\\
\langle \hat{Y}_{n}(t)\rangle=[r (t)]^n \sin(n \alpha).
\end{align}
\end{subequations}

To interpret the physical role of this set of operators, let us come back to the classical states given in Eqs. \eqref{eq:clasico1} and \eqref{eq:clasico2}. In this case, hypothesis (H2) implies that $r_n(t) > 0$, $\forall n$. Therefore, it is guaranteed that $\langle \hat{X}_n (t) \rangle > 0$, $\forall t$, if $\langle \hat{X}_n (0) \rangle > 0$; $\langle \hat{X}_n (t) \rangle < 0$, $\forall t$, if $\langle \hat{X}_n (0) \rangle < 0$, and $\langle \hat{X}_n (t) \rangle = 0$, $\forall t$, if $\langle \hat{X}_n (0) \rangle = 0$. And the same holds for $\hat{Y}_n$. 

From these facts, we obtain:

{\bf (C2).--} The following set of observables,
\begin{subequations}\label{eq:signs}
\begin{align}
&\hat{\mathcal{C}}_{n}\equiv \textrm{sign}\,(\hat{X}_{n}),\\
&\hat{\mathcal{K}}_{n}\equiv \textrm{sign}\,(\hat{Y}_{n}),
\end{align}
\end{subequations}
with
\begin{equation}
\textrm{sign} (x) = \begin{cases} \phantom{-}1 &\textrm{ if } x>0, \\ \phantom{-}0 &\textrm{ if } x=0, \\ -1 &\textrm{ if } x<0, \end{cases}
\end{equation}
are constants of motion in the symmetry-breaking for classical states. 

From all these facts, we formulate the key point of our theory, which consists in a generalization of (C1) and (C2) to arbitrary quantum states within a continuous symmetry-breaking phase.

{\em The operators $\hat{\mathcal{C}}_{n}$ and $\hat{\mathcal{K}}_{n}$, $\forall n \in \mathbb{N}$, defined in Eqs. \eqref{eq:signs}, are constants of motion within $\mathcal{H}_{SB}$ in any quantum system whose Hamiltonian is invariant under any rotation generated by $\hat{L}$.}

In Sec. \ref{sec:quantum} we will show that this is true for the model we have chosen to illustrate our theory. For the moment, we remark that this result requires the following technical assumption:

Let $\mathbb{P}_{\pm}(X_{n},t)$ and $\mathbb{P}_{\pm}(Y_{n},t)$ denote the probability of measuring positive or negative values of $\hat{X}_{n}$, $\hat{Y}_{n}$ at time $t$. Then, 
\begin{subequations}\label{eq:propertyprobability}
\begin{align}
&\mathbb{P}_{\pm}(X_{n},t)=\mathbb{P}_{\pm}(X_{n},0),\,\,\,\forall t\geq0,\\
&\mathbb{P}_{\pm}(Y_{n},t)=\mathbb{P}_{\pm}(Y_{n},0),\,\,\,\forall t\geq0.
\end{align}
\end{subequations}
Physically, this means that the observables $\hat{X}_{n}$ and $\hat{Y}_{n}$ behave as the typical order parameters of \textit{discrete} symmetry-breaking phase transitions: if, due to the symmetry breaking, the system can only exhibit positive (or negative) values of these observables at $t=0$, then this property holds true for all subsequent times. In other words, the symmetry cannot be restored. 

The constants of motion \eqref{eq:signs} are obtained directly from the order parameters $\{\hat{X}_{n},\hat{Y}_{n}\}$ as follows. Let $\ket{X_{n}^{m}}$ be an eigenstate of $\hat{X}_{n}$ with eigenvalue $X_{n}^{m}$; then, $\hatmath{C}_{n}\ket{X_{n}^{m}}=\ket{X_{n}^{m}}$, if $X_{n}^{m}>0$; $\hatmath{C}_{n}\ket{X_{n}^{m}}=-\ket{X_{n}^{m}}$, if $X_{n}^{m}<0$; and $\hatmath{C}_{n}\ket{X_{n}^{m}}=0$, if $X_{n}^{m}=0$. From the definitions \eqref{eq:signs}, it follows that $\hatmath{C}_{n}$ and $\hatmath{K}_{n}$ are discrete operators with three eigenvalues, $-1,0,1$, and their instantatenous expectation values in any state are restricted to $\langle\hatmath{C}_{n}(t)\rangle\in[-1,1]$, $\forall t$. 

From the preceding discussion it follows that any (non-trivial) function of all these observables: 
\beq
\hat{\rho}_{\textrm{SB}}=\hat{\rho}_{\textrm{SB}}(\hat{H},\hat{L},\{\hat{\mathcal{C}}_{n}\}_{n},\{\hat{\mathcal{K}}_{n}\}_{n}),\,\,\,n\in\mathbb{N},
\eeq
is a symmetry-breaking equilibrium state.

\subsection{Physical insight}\label{sec:insight}

The results of the previous section imply that the initial expectation values of $\hat{\mathcal{C}}_n$ and $\hat{\mathcal{K}}_n$ are conserved by the unitary time evolution, within $\mathcal{H}_{SB}$. However, the physical role of these operators is not yet so clear. In particular, we may wonder how many of them provide relevant information, and how many of them are irrelevant in practice. Here, we discuss two important consequences of the former framework, implying that the whole set of $\hat{\mathcal{C}}_n$ and $\hat{\mathcal{K}}_n$ play an important physical role.

As we have pointed out below, results in Sec. \ref{sec:constants} require that there exist equilibrium states that do not remain invariant under any rotation, that is, equilibrium states in which the continuous symmetry is totally broken. Let us now explore what happens if this is not true, that is, if we are in a phase in which all equilibrium states remain invariant under particular rotations. To do so, we choose a particular form of $\hat{O}$, which will be of relevance for the examples provided in the next sections of this work. We assume that $\hat{O}$ and $\hat{O}^{\dagger}$ act like raising and lowering operators for the quantum number defined by $\hat{L}$,
\begin{subequations}
\label{eq:propertyO}
\begin{align}
&\hat{O}\ket{E_{n}^{\ell}}=\sum_{m=1}^{D_{\ell+1}} c_{m}\ket{E_{m}^{\ell+1}},\,\,\forall \ell\in\mathbb{Z}, \\
&\hat{O}^{\dagger} \ket{E_{n}^{\ell}}=\sum_{m=1}^{D_{\ell-1}} c^*_{m}\ket{E_{m}^{\ell-1}},\,\,\forall \ell\in\mathbb{Z},
\end{align}
\end{subequations}
where, in general, the coefficients $c_m$ are not necessarily restricted to the thin spectrum. For some purposes, it also may be assumed that $\hat{O}$ and $\hat{O}^{\dagger}$ fulfill $[\hat{O},\hat{L}] = - \hat{O}$ and $[\hat{O}^{\dagger},\hat{L}] = \hat{O}^{\dagger}$, in order to build the complete set of generators of SU(2) from $\hat{O}$, $\hat{O}^{\dagger}$ and $\hat{L}$ (see, for example, \cite{Koma1994}, where this condition is used to prove that continuous symmetry-breaking is necessarily linked to the existence of a thin spectrum). We will see later that this holds true for the particular physical model that we have chosen to illustrate our theory. Notwithstanding, Eqs. \eqref{eq:propertyO} are enough for what follows:

$\bullet$ If $E_{m}^{2i}=E_{m}^{2j}$ and $E_{m}^{2i+1}=E_{m}^{2j+1}$ but $E_{m}^{2i} \neq E_{m}^{2i+1}$, then $\hat{\mathcal{C}}_{2n-1}$ and $\hat{\mathcal{K}}_{2n-1}$ cannot be constants of motion (see Appendix \ref{sec:appendixI} for a proof). This means that the symmetry $\hat{R}(\alpha)$ is broken, but not maximally broken since a rotation of $\alpha=\pi$ degrees leaves $\hat{\rho}_{\textrm{SB}}$ invariant. Indeed, observe that for an linear superposition $\ket{\psi_m}=\sum_{j\in\mathbb{Z}}c_{j}\ket{E_{m}^{2j}}$, we have that 
\beq 
\hat{R}(\alpha)\ket{\psi_m}=\sum_{j\in\mathbb{Z}}e^{2j\alpha i}\ket{E_{m}^{2j}}=\ket{\psi_m}\iff \alpha=\pi.
\eeq
Therefore, generic equilibrium density matrices consisting in a statistical mixture of these kind of states,
\begin{equation}
\hat{\rho}_{\textrm{SB}} = \sum_m p_m \ket{\psi_m} \bra{\psi_m},
\end{equation}
fulfill
\beq
\hat{R}(\pi)\hat{\rho}_{\textrm{SB}}\hat{R}^{\dagger}(\pi)=\hat{\rho}_{\textrm{SB}}.
\eeq
This means that $\hat{R}(\pi)$ is a $\mathbb{Z}_{2}$ discrete symmetry. 

$\bullet$ The previous argument can be generalized. Assume that $E_{m}^{ki}=E_{m}^{kj}$ and $E_{m}^{ki+p}=E_{m}^{kj+p}$ where $k=2,3,\ldots$, $i,j\in\mathbb{Z}$, and $p=1,2,\ldots,k-1$. In this case, the operators $\hat{\mathcal{C}}_{kn}$ and $\hat{\mathcal{K}}_{kn}$ are the only constants of motion for all $n\in\mathbb{N}$. By the same reasoning as above, any rotation of $\alpha=2\pi/k$ leaves the equilibrium state invariant,
\beq
\hat{R}\left (\frac{2\pi}{k}\right)\hat{\rho}_{\textrm{SB}}\hat{R}^{\dagger}\left(\frac{2\pi}{k}\right )=\hat{\rho}_{\textrm{SB}},\,\,k=2,3,\ldots,
\eeq
and the operator $\hat{R}(2\pi/k)$ behaves as a $\mathbb{Z}_{k}$ discrete symmetry. In the limit $k\to\infty$, the equilibrium state becomes invariant under any rotation, which leads to a symmetric $\hat{\rho}_{\textrm{S}}$. 

On the other hand, a necessary condition for $\hat{\mathcal{C}}_n$ and $\hat{\mathcal{K}}_n$ to be constants of motion $\forall n$ is that $E_m^i = E_m^j$ $\forall i,j \in \mathbb{Z}$. As this implies that a generic superposition of all these degenerate eigenstates, $\ket{\psi_m} = \sum_{j \in \mathbb{Z}} c_j \ket{E_n^j}$, is not invariant under any rotation $\hat{R}(\alpha)$, we conclude that there exists equilibrium states fulfilling $\hat{R}(\alpha) \hat{\rho}_{\textrm{SB}} \hat{R}^{\dagger} (\alpha) \neq \hat{\rho}_{\textrm{SB}}$ $\forall \alpha$, as stated in hypothesis (H1). As a consequence, the number of conserved quantities in the set $\left\lbrace \hat{\mathcal{C}}_1, \hat{\mathcal{K}}_1, \hat{\mathcal{C}}_2, \hat{\mathcal{K}}_2, \ldots \right\rbrace$ provides a way to establish whether the continuous symmetry is totally or partially broken in the corresponding phase. This general classification is useful for realistic physical systems. For example, the continuous symmetry breaking in the Bose-Hubbard model corresponds to the scenario in which all the operators $\hat{\mathcal{C}}_n$, $\hat{\mathcal{K}}_n$ are constants of motion in the ordered phase, because the single bosonic creator operator is a good order parameter \cite{Sachdev}. On the contrary, in the superconducting phase transition, symmetry breaking is linked to the conservation of $\hat{\mathcal{C}}_{2n}$, $\hat{\mathcal{K}}_{2n}$, since the Cooper pairs are composed by two electrons and therefore we need to create two electrons to get a good order parameter \cite{Tinkhambook}.

\begin{figure}[h!]
\begin{tabular}{c}
\hspace{-1cm}\includegraphics[width=0.45\textwidth]{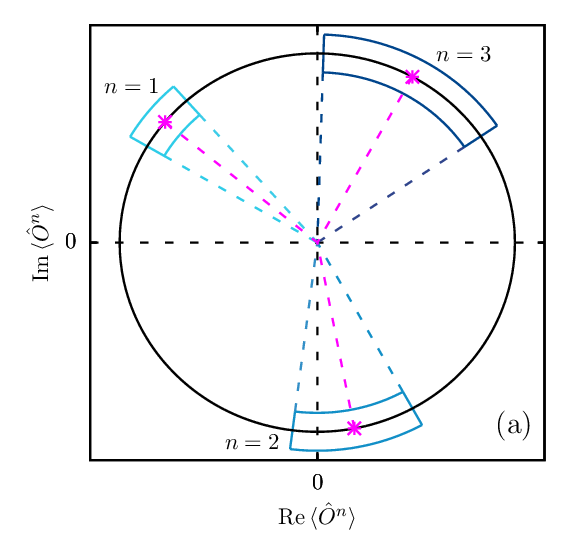}\\
\hspace{-1cm}\includegraphics[width=0.45\textwidth]{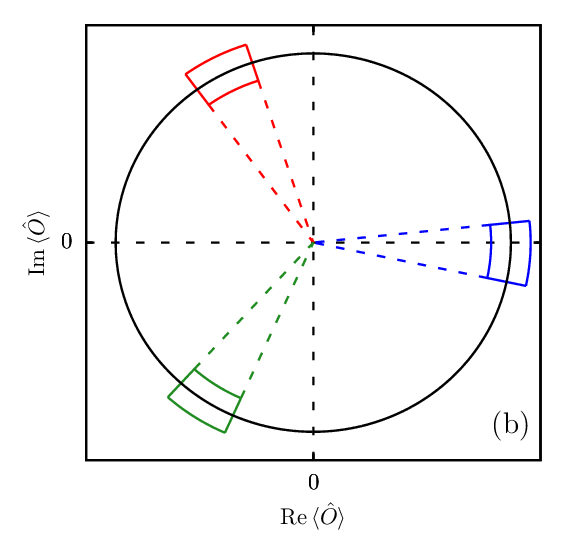}
\end{tabular}
\caption{Schematic representation of the (complex) expectation value of the order parameter $\hat{O}^{n}$ ($n=1,2,3$), see \eqref{eq:H2b}. (a) Expectation values for classical states, with no uncertainty in the symmetry-breaking angle (magenta points), and for quantum states with increasing uncertainty in the angle $\alpha$ (light blue, blue and dark blue annulus sectors). (b) Expectation value of $\hat{O}$ for a quantum superposition of three states defined by different angles. }
\label{fig:schematic}
\end{figure}

The second consequence is that the operators $\hat{\mathcal{C}}_n$ and $\hat{\mathcal{K}}_n$ allow us to distinguish between classical and quantum symmetry-breaking states. To understand why, let us come back to the classical states, defined in Eqs. \eqref{eq:clasico1} and \eqref{eq:clasico2}. From these equations we immediately obtain,
\begin{subequations}
\label{eq:classical}
\begin{align}
& \hat{X}_n \ket{\alpha} = r^n \cos(n \alpha) \ket{\alpha}, \\
& \hat{Y}_n \ket{\alpha} = r^n \sin(n \alpha) \ket{\alpha}.
\end{align}
\end{subequations}

From here, we can define a quantum symmetry-breaking state as an arbitrary \textit{superposition of the classical states},
\begin{equation}
\label{eq:quantum}
\ket{\Psi} = \sum_{\alpha\in\Lambda} c_{\alpha} \ket{\alpha},
\end{equation}
where $\Lambda$ may be either a discrete or a continuous set, and $\sum_{\alpha} | c_{\alpha}|^2=1$. If $\Lambda$ is a discrete set, $|c_{\alpha}|^2$ represents the probability of observing the system in the angle $\alpha$ in a measurement; if $\Lambda$ is a continuous set, $|c_{\alpha}|^2 \equiv \rho_{\alpha}$ represents the corresponding probability density.

From Eqs. \eqref{eq:classical} it is obvious that a classical state is also an eigenstate of $\hat{\mathcal{C}}_n$ and $\hat{\mathcal{K}}_n$:
\begin{subequations}
\label{eq:classical2}
\begin{align}
& \hat{\mathcal{C}}_n \ket{\alpha} = \textrm{sign} [\cos(n \alpha)] \ket{\alpha}, \\
& \hat{\mathcal{K}}_n \ket{\alpha} = \textrm{sign} [\sin(n \alpha)] \ket{\alpha}.
\end{align}
\end{subequations}
As a consequence, the expectation values of these operators in a classical state can be either (i) $\langle \hatmath{C}_{n}\rangle=\pm 1$ and $\langle \hatmath{K}_{n}\rangle=\pm 1$ for all $n\in\mathbb{N}$, or (ii) $\langle\hatmath{C}_{n}\rangle=0$ and $\langle\hatmath{K}_{n}\rangle=\pm 1$ for certain $n$ such that $n \alpha = k \pi/2$, with $k \in \mathbb{Z}$, or viceversa if $n\alpha=k\pi$. 

On the contrary, from Eq. \eqref{eq:quantum}, it follows that\footnote{This implies assuming that $\bra{\beta} \ket{\alpha} = \delta_{\beta \alpha}$ $\forall \alpha,\beta \in [0,2\pi)$. This cannot happen in any set of real quantum states; as an example, coherent states (the real quantum states closest to the classical ones) are not orthonormal. We remind here that the states defined in Eqs. \eqref{eq:clasico1} and \eqref{eq:clasico2} are fictitious. Notwithstanding, we will see in Sec. \ref{sec:quantum} that they can be asymptotically approached in a fully-connected model, for which the classical and the thermodynamic limit coincide.}
\begin{subequations}
\label{eq:quantum2}
\begin{align}
&\bra{\Psi} \hat{\mathcal{C}}_n \ket{\Psi} = \sum_{\alpha} |c_{\alpha}|^2  \textrm{sign}[\cos(\alpha n)], \\
&\bra{\Psi} \hat{\mathcal{K}}_n \ket{\Psi} = \sum_{\alpha} |c_{\alpha}|^2  \textrm{sign}[\sin(\alpha n)].
\end{align}
\end{subequations}
This means that, in general, we can obtain any values $\langle \hat{\mathcal{C}}_n \rangle \in [-1,1]$ and $\langle \hat{\mathcal{K}}_n \rangle \in [-1,1]$ for a quantum state, consisting in a superposition of classical states with well-defined symmetry-breaking directions. Therefore, {\em it suffices to obtain expectation values for $\hatmath{C}_n$ and/or $\hatmath{K}_n$ to safely conclude that the symmetry-breaking direction $\alpha$ is not well-defined, but it has quantum fluctuations}. In such a case, we can define a probability density, $\rho(\alpha)$, for the observation of a particular privileged direction $\alpha$ in a measurement, as we will see in Sec. \ref{sec:quantum}; the larger the number of expectation values $\langle \hat{\mathcal{C}}_n\rangle$ and $\langle \hat{\mathcal{K}}_n\rangle$ known, the more information about it. It also follows from \eqref{eq:quantum2} that a perfectly symmetric state in $\mathcal{H}_{SB}$, for which $\rho(\alpha)=1/(2 \pi)$, $\forall \alpha \in [0, 2 \pi)$, is characterized by $\langle \hat{\mathcal{C}}_n \rangle=\langle \hat{\mathcal{K}}_n \rangle=0$, $\forall n$, with all of these operators being constants of motion. However, in $\mathcal{H}_{S}$ we have that $\overline{\langle \hat{\mathcal{C}}_n \rangle}= \overline{\langle \hat{\mathcal{K}}_n \rangle}=0$, where $\overline{\bullet}$ represents an infinite-time average, but none of the $\hat{\mathcal{C}}_n$ and $\hat{\mathcal{K}}_n$ operators are constant in this case. 

We provide two schematic illustrations of these facts in Fig. \ref{fig:schematic}.  In Fig. \ref{fig:schematic}(a), we illustrate the behavior of a classical state, represented by magenta stars. We show it in the complex plane for $n=1$, $2$ and $3$; in all cases, the classical state is represented by a single point. It is characterized by $\langle \hat{\mathcal{C}}_1 \rangle=\langle \hat{\mathcal{K}}_2 \rangle=-1$ and $\langle \hat{\mathcal{K}}_1 \rangle=\langle \hat{\mathcal{C}}_2 \rangle= \langle \hat{\mathcal{C}}_3 \rangle=\langle \hat{\mathcal{K}}_3 \rangle=1$. In the same panel, we represent a quantum case consisting in the superposition with a flat probability distribution on a finite support in the second sector of the circumference. Due to the shape of Eqs. \eqref{eq:quantum2}, the behavior of the constants of motion with $n=2$ can be inferred from a flat distribution with a support two times wider, and we can reason in the same way for $n=3$. As a consequence, we obtain $\langle \hat{\mathcal{C}}_1 \rangle=\langle \hat{\mathcal{K}}_2 \rangle=-1$ and $\langle \hat{\mathcal{K}}_1 \rangle= \langle \hat{\mathcal{C}}_3 \rangle=\langle \hat{\mathcal{K}}_3 \rangle=1$, as in the classical case, but with $0 < \langle \hat{\mathcal{C}}_2 \rangle < 1$, which cannot be observed with a classical symmetry-breaking state. In Fig. \ref{fig:schematic}(b) we show a quantum state whose probability distribution on $\alpha$ is the superposition of three different flat distributions with finite supports, giving rise to a more complex pattern for the expectation values of $\hat{\mathcal{C}}_n$ and $\hat{\mathcal{K}}_n$.

\section{Model Hamiltonian}\label{sec:modelham}

The core of our theory for continuous symmetry-broken phases has been laid out in the preceding part of our paper. Because our arguments are general, they do not only apply to a specific system. Our next goal is to test our theory in a toy-model for continuous symmetry-breaking, which we present in this section.

We consider the two-dimensional limit of the vibron model, which has played an important role in the description of dynamical properties of certain molecules \cite{Khalouf2021,PerezBernal2005,Iachello2003,Iachello1996}. Its structure has been studied in detail in, e.g., \cite{PerezBernal2008,Khalouf2022,PerezBernal2010}. This is a two-level interacting bosonic model where the lower level is populated by $\hat{\sigma}$ bosons and the upper level contains $\hat{\tau}_{+}$ and $\hat{\tau}_{-}$ bosons. All of them satisfy the standard boson commutation relations, i.e., $[\hat{\sigma},\hat{\sigma}^{+}]=1$, $[\hat{\tau}_{\alpha},\hat{\tau}_{\alpha'}^{\dagger}]=\delta_{\alpha\alpha'}$, $\alpha,\alpha'\in\{+,-\}$.  The Hamiltonian reads
\beq\label{eq:ham}
\hat{H}=(1-\xi)\hat{n}_{\tau}+\frac{\xi}{N-1}\hat{P},\,\,\xi\in[0,1],
\eeq
with $\hat{n}_{\tau}=\hat{n}_{+}+\hat{n}_{-}$ being the number of bosons in the upper level, $\hat{n}_{\pm}=\hat{\tau}_{\pm}^{\dagger}\hat{\tau}_{\pm}$, and $\hat{P}=\hat{N}^{2}-\hat{L}^{2}-\frac{1}{2}(\hat{D}_{+}\hat{D}_{-}+\hat{D}_{-}\hat{D}_{+})$ is the so-called pairing operator, with $\hat{D}_{+}=\sqrt{2}(\hat{\tau}_{+}^{\dagger}\hat{\sigma}-\hat{\sigma}^{\dagger}\hat{\tau}_{-})$ and $\hat{D}_{-}=\hat{D}_{+}^{\dagger}$. The total bosonic operator $\hat{N}=\hat{n}_{\sigma}+\hat{n}_{+}+\hat{n}_{-}$ is a conserved quantity. The Hamiltonian \eqref{eq:ham} is invariant under the rotation $\hat{R}(\alpha)=e^{i\alpha\hat{L}}$, for all $\alpha\in[0,2\pi)$,
which is a continuous symmetry. Here,  $\hat{L}= \hat{n}_{+}-\hat{n}_{-}$ is the angular momentum operator. The system has two degrees of freedom and is integrable since $\{\hat{H},\hat{L}\}$ is a set of two independent integrals of motion. Observe that $\hat{H}$ is an even function of $\hat{L}$, and therefore $E_{n}^{\ell}=E_{n}^{-\ell}$. The dimension $D_{\ell}$ of each $\ell$ sector depends on the parity of $N$ and $\ell$. If $N-\ell\equiv 0\,\textrm{mod}\,2$, then $D_{\ell}=(N-|\ell|)/2+1$, whereas if $N-\ell\equiv 1\,\textrm{mod}\,2$, then $D_{\ell}=(N-|\ell|+1)/2$. The dimension of the finite-$N$ Hamiltonian matrix is $D=\sum_{\ell} D_{\ell}=(N+1)(N+2)/2$. Therefore, the infinite-size limit corresponds to $N\to\infty$.

The model exhibits a second-order QPT at $\xi_{c}=1/5$, where $\textrm{d}^{2}\epsilon_{\textrm{GS}}/\textrm{d}\xi^{2}$ becomes discontinuous in the infinite-size limit. The ground-state energy bifurcates as a function of the control parameter $\xi$ and its value reads 

\beq \label{eq:Egs}\epsilon_{\textrm{GS}}=
    \begin{cases}
      \xi & \text{if $0\leq \xi\leq \xi_{c},$}\\
      \frac{-1+10\xi-9\xi^{2}}{16\xi} & \text{if $\xi_{c}\leq \xi\leq 1.$}
    \end{cases}
\eeq

 Here, the energy per particle is $\epsilon\equiv E/N$, where $E$ denotes the actual (extensive) eigenvalues of the Hamiltonian \eqref{eq:ham}. 

For $\xi\geq \xi_{c}$, the the discontinuity in the ground-state manifold gets transferred onto high-lying eigenvalues, leading to an excited-state quantum phase transition (ESQPT) \cite{Caprio2008,Cejnar2021} at a critical energy \cite{PerezBernal2008} given by
\beq\label{eq:Eesqpt}
\epsilon_{c}=\xi,\,\,\,\xi\geq \xi_{c}.
\eeq
For our purposes, it will prove useful to consider an extension of the Hamiltonian above, 
\beq\label{eq:Hmu}
\hat{H}_{\mu}=\hat{H}+\mu\hat{L},\,\,\mu\in\mathbb{R}.
\eeq
The additional term $\mu\hat{L}$ has no effect on the symmetry properties discussed above, but it does have an impact on eigenlevel degeneracies (see below).

Because \eqref{eq:ham} describes a fully-connected model, the $N\to\infty$ limit coincides with the classical limit, $\hbar_{\textrm{eff}}\to0$, which we briefly describe in the following section. Thus, we can expect that the classical states $\ket{\alpha}$ defined in Eqs. \eqref{eq:clasico1} and \eqref{eq:clasico2} to be the $N \rightarrow \infty$ limit of coherent states (see below).

\section{Classical limit}\label{sec:classicallimit}
In the classical limit, a quantum operator is replaced by a dynamical function defined on a subset of $\mathbb{R}^{2f}$, where $f$ is the number of classical degrees of freedom, $\hat{O}\to O(\{q_{k},p_{k}\}_{k=1}^{f})$. To obtain the classical limit, we employ the Holstein-Primakoff transformation \cite{Holstein1940}, whose details are in Appendix \ref{sec:appendixII}. We consider the coherent state 
\beq\label{eq:coherent}
\ket{\Phi}=\frac{1}{\mathcal{N}}\sum_{m=0}^{N}\frac{1}{m!}\sqrt{\frac{N^{m}}{(N-m)!}}(\hat{\sigma}^{\dagger})^{N-m}(\alpha_{+}\hat{\tau}_{+}^{\dagger}+\alpha_{-}\hat{\tau}_{-}^{\dagger})^{m}\ket{0},
\eeq
where $\mathcal{N}$ is such that $\bra{\Phi}\ket{\Phi}=1$, and $\alpha_{\pm}\equiv\alpha_{\pm}(\mathbf{q},\mathbf{p})\in\mathbb{C}$ are related to the canonical position and momentum as 
\begin{subequations}\label{eq:alphamasmenos}\begin{align}
&\alpha_{+}(\mathbf{q},\mathbf{p})=-\frac{1}{2}(q_{1}+p_{2})+\frac{i}{2}(q_{2}-p_{1}),\\
&\alpha_{-}(\mathbf{q},\mathbf{p})=\frac{1}{2}(q_{1}-p_{2})+\frac{i}{2}(p_{1}+q_{2}).\end{align}
\end{subequations}
The circular boson operators $\hat{\tau}_{\pm}$ entering the Hamiltonian \eqref{eq:ham} can be mapped onto cartesian boson operators $\hat{\tau}_{1,2}$ according to \cite{PerezBernal2008}
\beq\label{eq:taumasmenos}
\hat{\tau}_{\pm}=\mp \frac{\hat{\tau}_{1}\mp i\hat{\tau}_{2}}{\sqrt{2}}.
\eeq
In the classical limit, each cartesian boson is substituted for pair of coordinate-momentum operators, 
\beq\label{eq:tau12}
\hat{\tau}_{k}=\sqrt{\frac{N}{2}}(\hat{q}_{k}+i\hat{p}_{k}),\,\,k=1,2.
\eeq
These operators satisfy
\beq\label{eq:commutatorqp}
[\hat{q}_{k},\hat{p}_{k'}]=\frac{i}{N}\delta_{kk'}\equiv i\hbar_{\textrm{eff}}\delta_{kk'}.
\eeq
Observe that the $N\to\infty$ limit is mathematically identical to the limit $\hbar_{\textrm{eff}}=1/N\to0$, and in this limit the $\hat{q}_{k}$ and $\hat{p}_{k}$ operators behave as scalars, $\hat{q}_{k}\to q_{k}$. Importantly, these are pairs of canonically conjugated variables, $\{q_{k},p_{k}\}=1$, which makes the study of dynamics a trivial matter from the Hamilton equations. In three-dimensional space, $q_{1,2}$ and $p_{1,2}$ represent the canonical position and momentum along the $x$ and $y$ axes.

The classical analogue of the Hamiltonian \eqref{eq:Hmu} is thus given by the $N-$independent function
\beq\label{eq:classicalham}
\begin{split}
H_{\mu}(\mathbf{q},\mathbf{p})&=\frac{\bra{\Phi}\hat{H}_{\mu}\ket{\Phi}}{N}=\frac{1}{2}(1-\xi)\Omega^{2}\\
&+\xi\left[1-(q_{1}^{2}+q_{2}^{2})(2-\Omega^{2})-(p_{1}q_{2}-p_{2}q_{1})^{2}\right]\\&+\mu(p_{2}q_{1}-p_{1}q_{2}),
\end{split}
\eeq 
where $\Omega^{2}\equiv q_{1}^{2}+p_{1}^{2}+q_{2}^{2}+p_{2}^{2}\leq 2$. The classical phase space is a four-dimensional hypersphere centered at $(\mathbf{0},\mathbf{0})$, i.e, $\mathcal{M}=\{(\mathbf{q},\mathbf{p})\in\mathbb{R}^{4}: \Omega^{2}\leq 2\}$. 

\begin{figure*}[t]
\begin{tabular}{c c}
\hspace{-1.5cm}\includegraphics[width=0.4\textwidth]{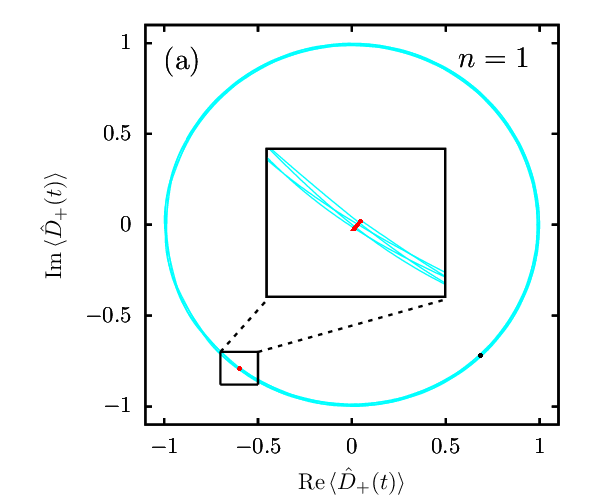} &\hspace{-0.9cm}\includegraphics[width=0.4\textwidth]{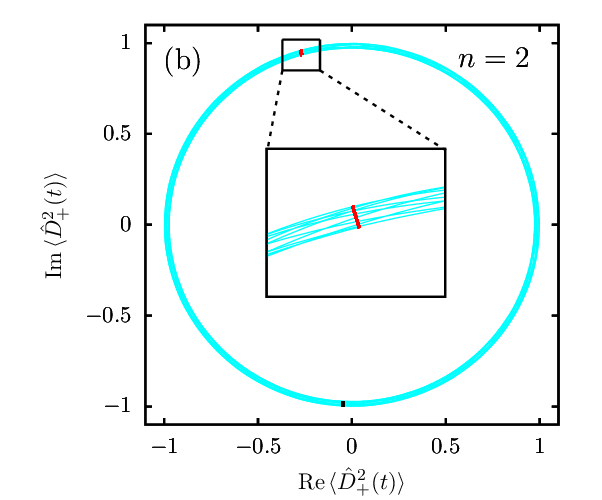}\\
\hspace{-1.5cm}\includegraphics[width=0.4\textwidth]{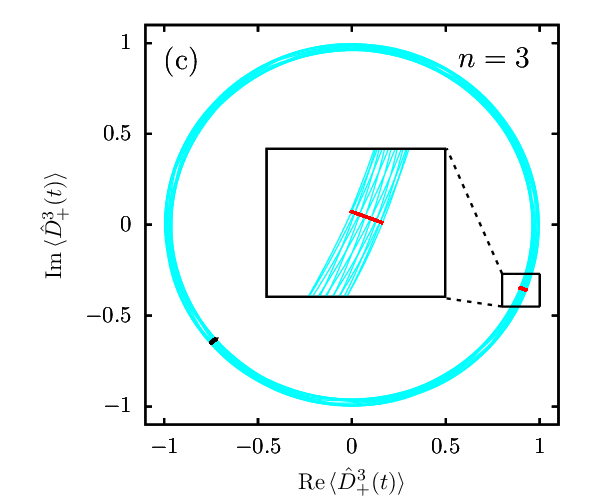} &\hspace{-0.9cm}\includegraphics[width=0.4\textwidth]{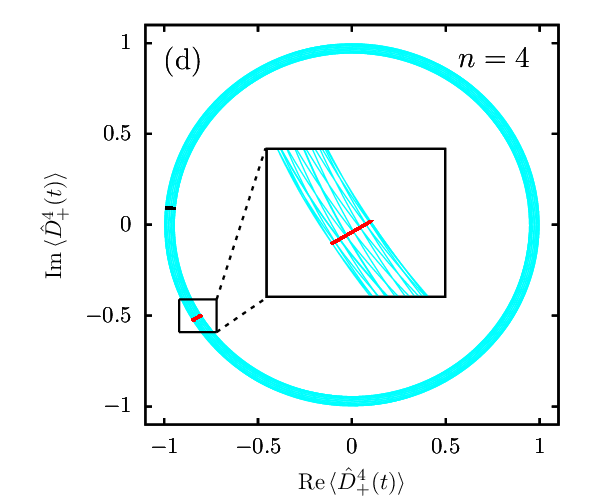}

\end{tabular}
\caption{Classical dynamics of the $\hat{D}^{n}_{+}$ operator obtained from solving the Hamilton equations \eqref{eq:hameq} at $\xi=0.8$. The dynamics take place in the complex plane $(\Re\langle \hat{D}_{+}^{n}(t)\rangle,\Im\langle\hat{D}_{+}^{n}(t)\rangle)$. Different panels depict the evolution of several powers of the operator: (a) $n=1$, (b) $n=2$, (c) $n=3$ and (d) $n=4$. We consider two initial states at $\xi=0.6$ as the ground-state coordinates $(q_{1},q_{2})$ ($p_{1}=0,p_{2}=0$) satisfying $q_{1}^{2}+q_{2}^{2}=R^{2}$ in Eq. \eqref{eq:R}, with $q_{1}=0.55$ (red, cyan), 
and $q_{1}=-0.63$ (black). Black and red trajectories are obtained from the classical Hamiltonian \eqref{eq:classicalham} with $(\xi=0.8$, $\mu=0)$, and cyan trajectories correspond to $(\xi=0.8,\mu=0.3)$. In (a), the black trajectory is defined by an angle $\alpha\approx 1.742\pi$, and the red trajectory has $\alpha\approx1.294\pi$. The energy of all trajectories is $\epsilon\approx 0.106$. }
\label{fig:Dmasclassical}
\end{figure*}

The ground-state of the classical system is obtained through minimization of \eqref{eq:classicalham}. For $\mu=0$, the result is $\mathbf{p}_{\textrm{GS}}=(0,0)$, and $\norm{\mathbf{q}_{\textrm{GS}}}^{2}=R^{2}$ with 
\beq\label{eq:R} R^{2}(\xi)=
    \begin{cases}
      0 & \text{if $0\leq \xi\leq \xi_{c},$}\\
      \frac{5\xi -1}{4\xi} & \text{if $\xi_{c}\leq \xi\leq 1.$}
    \end{cases} 
\eeq
Therefore, the ground-state energy, $\epsilon_{\textrm{GS}}=H_{\mu=0}(\mathbf{q}_{\textrm{GS}},\mathbf{p}_{\textrm{GS}})$, coincides with \eqref{eq:Egs}. Observe that for $0\leq \xi\leq \xi_{c}$, the ground-state is unique, but for $\xi_{c}\leq \xi\leq 1$ there are infinitely many configurations of $(q_{1},q_{2})$ that yield the same energy: the ground-state manifold is continuously symmetry-broken. In the complex plane, the coordinates $(q_{1},q_{2})$ can be used to represent a complex number $z=q_{1}+iq_{2}=|z|e^{i\alpha}$, with a well-defined argument, $\alpha=\textrm{Arg}(z)$, and with a fixed modulus, $|z(\xi)|=R(\xi)>0$. The only difference in all configurations belonging to the ground-state manifold lies in the angle $\alpha$. For this reason, the $\hat{D}_{+}$ operator, which acts as a raising ladder operator changing the $\ell$ quantum number as $\ell\to\ell+ 1$ while keeping fixed the total number of bosons, $N$, can be used as an the order parameter $\hat{O}$ in \eqref{eq:H2b}: indeed, in the classical limit, this operator is described by the function
\beq \label{eq:Dmasclassical}
D_{+}(\mathbf{q},\mathbf{p})=\frac{\bra{\Phi}\hat{D}_{+}\ket{\Phi}}{N}=-\sqrt{2-\Omega^{2}}(q_{1}+iq_{2}),\eeq
where $\ket{\Phi}$ is the coherent state \eqref{eq:coherent} (see Appendix \ref{sec:appendixII} for details). Each pair $(q_{1},q_{2})$ corresponds to a single angle in the polar representation of $D_{+}$. 

Finally, for $\xi\geq \xi_{c}$, the origin $(\mathbf{0},\mathbf{0})$ becomes an unstable stationary point, leading to an ESQPT defined in \eqref{eq:Eesqpt}. 

\subsection{Classical dynamics}\label{sec:classicaldynamics}

In this section we investigate the dynamics of our model \eqref{eq:Hmu}, whose behavior in the $N\to\infty$ limit is described by the classical Hamiltonian \eqref{eq:classicalham}. 

Classical dynamics is governed by the Hamilton equations:
\beq\label{eq:hameq}
\frac{\textrm{d}\mathbf{q}}{\textrm{d}t}=\frac{\partial H(\mathbf{q},\mathbf{p})}{\partial \mathbf{p}},\,\,\,\frac{\textrm{d}\mathbf{p}}{\textrm{d}t}=-\frac{\partial H(\mathbf{q},\mathbf{p})}{\partial \mathbf{q}}.
\eeq
We perform classical quenches, whereby an initial state $(\mathbf{q},\mathbf{p})\in\mathcal{M}$ is chosen in the initial Hamiltonian $H(\xi_{i})$ and is then left to evolve under $H(\xi_{f})$ (therefore, the system of differential equations are derived from $H(\xi_{f})$). The initial state can be chosen to correspond to the ground-state of $H(\xi_{i})$, for example. Since all quantum operators become dynamical functions in the classical limit, $\hat{O}\to O(\mathbf{q},\mathbf{p})$, all one needs to do is solve the system of differential equations to obtain $(\mathbf{q}(t),\mathbf{p}(t))$ to calculate the classical evolution $O(\mathbf{q}(t),\mathbf{p}(t))$. We are interested in the dynamics of the order parameter $\hat{O}=\hat{D}_{+}$. (qualitatively analogous results are obtained with the $\hat{D}_{-}$ operator). Observe that because $\hat{D}_{+}$ is not a physical observable ($\hat{D}_{+}^{\dagger}\neq \hat{D}_{+}$), the dynamical function \eqref{eq:Dmasclassical} is complex-valued. Thus, its evolution may be represented in the complex plane. 

We are interested in two kinds of classical quenches: $(\xi=0.6,\mu=0)\to(\xi=0.8,\mu=0)$ and $(\xi=0.6,\mu=0)\to(\xi=0.8,\mu=0.3)$. 
In both cases, the initial state ($\mathbf{q}(0),\mathbf{p}(0)$) is given by a certain configuration of the ground-state manifold at $\xi=0.6$; as we have mentioned before, these belong to a circumference of radius $R(\xi)$, $q_{1}^{2}+q_{2}^{2}=R^{2}(\xi)$, given by \eqref{eq:R}. We fix $q_{1}$ and obtain the remaining coordinate as $q_{2}=+\sqrt{R^{2}(\xi)-q_{1}^{2}}$. 

The dynamical function $\hat{D}^{n}_{+}(t)$, with $n=1,2,3,4$, is represented in Fig. \ref{fig:Dmasclassical}. In Fig. \ref{fig:Dmasclassical}(a) we focus on $D_{+}(t)$. For the quench with $\mu=0\to\mu=0$, the red trajectory corresponds to an initial condition with $q_{1}=0.55$, and the black trajectory has $q_{1}=-0.63$, while in the case of the $\mu=0\to\mu=0.3$ quench we only represent a cyan trajectory with $q_{1}=0.55$. The main point is that the dynamics satisfy our hypotheses (H1) and (H2) and, in particular, \eqref{eq:H2} with $n=1$: $D_{+}(t)=r_{1}(t)e^{i\alpha}$ with $\alpha$ constant. This immediately implies that the sign of the expectation value of the $\hat{X}_{1}$ and $\hat{Y}_{1}$ operators in \eqref{eq:XnYn} is completely determined its the initial condition. The state is therefore symmetry-breaking, as $D_{+}(t)\neq 0$. The situation changes completely for the quench $\mu=0\to\mu=0.3$: the angle $\alpha$ is not constant anymore, and the constancy of the sign of the expectation value of $\hat{X}_{1}$ and $\hat{Y}_{1}$ does not necessarily follow. This state is symmetric, with an infinite-time average $\overline{D_{+}}=\lim_{T\to\infty}\frac{1}{T}\int_{0}^{T}D_{+}(t)\textrm{d}t=0$. Figs. \ref{fig:Dmasclassical}(b-d), which show the behavior of $D_{+}(t)$ raised to increasing powers, $D_{+}^{n}(t)$, provide additional corroboration of this result: $D_{+}^{n}(t)=r_{n}(t)e^{in\alpha}$ with $\alpha$ constant implies the constancy of the sign of the expectation value of $\hat{X}_{n}$ and $\hat{Y}_{n}$ for all $n\in\mathbb{N}$. The rotation of an $n\alpha$ angle is clear from the dynamics. Also, it can be observed that as $n$ increases the fluctuations in $r_{n}(t)$ increase too: the dynamics is restricted to an annulus of increasing section.

\section{Quantum results}\label{sec:quantum}

In the previous section we have shown that our model fulfills hypotheses (H1) and (H2), and therefore it also fulfills the consequences (C1) and (C2) in its classical limit. We test now if the behavior of its quantum version also agrees with our theory. 

\subsection{Spectral properties}\label{sec:spectralproperties}

\begin{figure*}[t!]
\begin{tabular}{c c}
\hspace{-1.5cm}\includegraphics[width=0.55\textwidth]{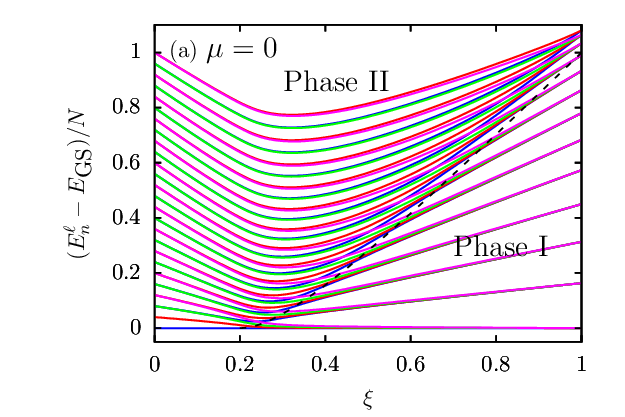} &\hspace{-1.5cm}\includegraphics[width=0.55\textwidth]{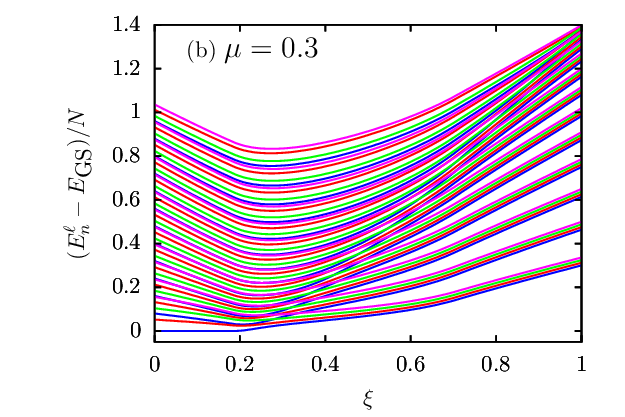}
\end{tabular}
\caption{Level flow diagram of the Hamiltonian Eq. \eqref{eq:ham} with (a) $\mu=0$ and (b) $\mu=0.3$. Eigenvalues are represented with respect to the ground-state energy (excitation spectrum). System size is $N=25$. Color lines represent eigenvalues with different $\ell$ quantum number: $\ell=0$ (blue), $1$ (red), $2$ (green), and $3$ (magenta). The dashed line in (a) represents the ESQPT critical energy in the $N\to\infty$ limit. }
\label{fig:levelflow}
\end{figure*}

The energy eigenvalues are one of the most important ingredients entering the time evolution generated by a time-independent Hamiltonian such as \eqref{eq:Hmu}. In Fig. \ref{fig:levelflow} we have represented the level flow diagram of the quantum Hamiltonian with $\xi\in[0,1]$ and (a) $\mu=0$ and (b) $\mu=0.3$. The eigenvalues are shown on the scale of the ground-state energy (excitation spectrum), and different colors are associated with different $\ell=0,1,2,3$ quantum numbers. In Fig. \ref{fig:levelflow}(a), we observe a clear separation of two phases characterized by markedly different spectral properties: phases I and II. The boundary between these phases is demarcated by the ESQPT separatrix given in \eqref{eq:Eesqpt}. Because the system size is only $N=25$, finite-size effects prevent the infinite-$N$ ESQPT critical energy from perfectly describing the phase separation. In any case, we observe that in phase I the eigenlevels $E_{n}^{\ell}$ and $E_{m}^{\ell'}$ seem to be degenerate if $n=m$ and $\ell,\ell'\in\mathbb{Z}$; however, in phase II we have that the $\ell$ quantum number of degenerate eigenlevels necessarily have the same parity ($\ell,\ell'=0,2,4,\ldots$ or $\ell,\ell'=1,3,5,\ldots$). This means that the continuous symmetry \eqref{eq:Rsym} is completely broken in phase I, as equilibrium states of the form $\ket{\psi}=\sum_{\ell\in\mathbb{Z}}c_{\ell}\ket{E_{n}^{\ell}}$ are only left invariant under the trivial rotation, $\hat{R}(\alpha=2\pi)$ [observe that $\hat{U}(t)\ket{\psi}=e^{-i\phi t}\ket{\psi}$ with $\phi=E_{n}^{0}=E_{n}^{1}=\ldots$, and therefore $\langle\hat{O}(t)\rangle=\langle \hat{O}(0)\rangle$ for any operator $\hat{O}$]. Yet, the $\hat{R}(\alpha)$ symmetry is only partially broken in phase II, as equilibrium states of the form $\ket{\psi}=\sum_{\ell\in\mathbb{Z}}c_{\ell}\ket{E_{n}^{2\ell}}$ are invariant under $\hat{R}(\alpha=\pi)$. The situation changes completely when the $\mu\hat{L}$ term is added to the Hamiltonian, as shown in Fig. \ref{fig:levelflow}(b). Although eigenlevel clustering seems to still take place at a given excitation energy, which hints at an ESQPT, the finite-$N$ Hamiltonian apparently loses the degeneracy properties of $\mu=0$. Instead, no degeneracies seem to exist, meaning that the continuous symmetry is not broken in this case. In finite-$N$ systems, the energy gap gradually increases as the ESQPT critical energy is approached. We will see later that these differences between the $\mu=0$ and $\mu=0.3$ cases entail crucial dynamical consequences. It is also worth pointing out that in phase I the lowest energy eigenvalues are the most degenerate, while in phase II it is the highest energy eigenvalues which show the smallest gap.

\begin{figure}[h!]
\hspace*{-1cm}\includegraphics[width=0.6\textwidth]{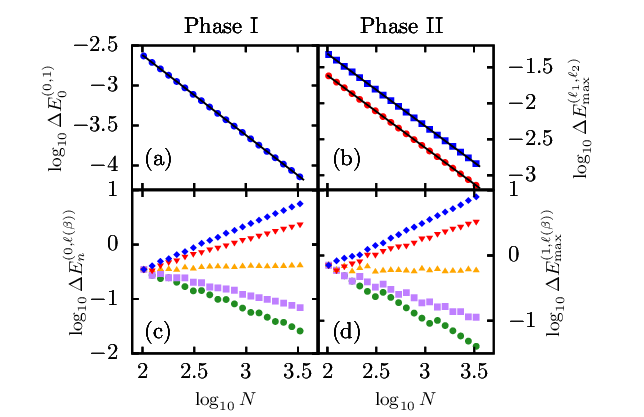}
\caption{Scaling of level degeneracies in phases I (left column) and II (right column) of the Hamiltonian Eq. \eqref{eq:Hmu} with $\mu=0$ and $\xi=0.6$. (a)  Scaling of distance of lowest energy eigenvalues in the $\ell=0,1$ sectors (blue squares), $\Delta E_{0}^{(0,1)}\propto N^{-1.0015}$. (b) Scaling of distance of maximum eigenvalues in the $\ell=1,3$ (blue squares), $\Delta E_{\max}^{(1,3)}\propto N^{-1.0045}$, and $\ell=0,2$ (red circles), $\Delta E_{\max}^{(0,2)}\propto N^{-1.0058}$, sectors. (c-d) Scaling of \eqref{eq:deltaE0elebeta} with, from bottom to top, $\beta=1/8,1/4,1/2,3/4,7/8$, and $\ell_{0}=10$, $N_{0}=103$.  }
\label{fig:paneldeg}
\end{figure}

The level flow diagram is a very convenient representation of the spectrum of a quantum system as a function of a control parameter. However, it does not provide any information about its infinite-size behavior, as $N$ is fixed. To corroborate the finite-$N$ impressions from Fig. \ref{fig:levelflow}(a)  ($\mu=0$), we study the scaling of the distance between apparently degenerate eigenlevels in Fig. \ref{fig:paneldeg}. The control parameter is fixed at $\xi=0.6$ for this analysis, similar results having been found for other values of $\xi$.  We analyze the following quantity energy gap as a function of $N$:
\beq\label{eq:deltaEl1l2}
\Delta E_{n}^{(\ell_{1},\ell_{2})}\equiv |E_{n}^{\ell_{1}}-E_{n}^{\ell_{2}}|,
\eeq
where $\ell_{1},\ell_{2}\in\mathbb{Z}$, $n=0,1,\ldots$.

Fig. \ref{fig:paneldeg}(a) shows the scaling of the distance of the lowest energy levels ($n=0$) with $\ell=0,1$ of phase I [c.f. Fig. \ref{fig:levelflow}(a), blue and red lines]. This energy gap vanishes as a power-law, $\Delta E_{0}^{(0,1)}\propto N^{-a}$, $a>0$. Fig. \ref{fig:paneldeg}(b) depicts a similar behavior in phase II for the highest energy eigenlevels with $\ell=0,2$ and $\ell=1,3$ [c.f. Fig. \ref{fig:levelflow}(a), blue-green and red-magenta lines]. It would therefore seem that in both phases bands of infinitely many degenerate eigenlevels appear only in the infinite-size limit, this behavior being typical of $\mathbb{Z}_{2}$ symmetry-broken phases \cite{Khalouf2023,Corps2021}. While our numerical analyses suggest this is the case, the situation is a bit more complicated than that, as we show below. 

In Fig. \ref{fig:levelflow}(c,d) we estimate the width of the degenerate bands in both phases I and II, respectively. We define the $d$-dependent $\ell$ quantum number $\ell(d)\equiv [\ell_{0}(N/N_{0})^{d}]$, where $\ell_{0}$ is a constant and $N_{0}$ is a fixed system-size of reference. Here, $[x]\in\mathbb{Z}$ denotes the closest integer number to $x$. Observe that $\ell(d)=1$, $\forall d$, if $N=N_{0}$, but otherwise $\ell(d)$ grows as a power-law. We are interested in 
\beq\label{eq:deltaE0elebeta}
\Delta E_{n^{*}}^{(0,\ell(d))}\equiv |E_{n^{*}}^{0}-E_{n^{*}}^{\ell(d)}|,
\eeq
where $n^{*}$ is a fixed band index. What is the behavior of this quantity as a function of $N$ for different $d$? In Fig. \ref{fig:paneldeg}(c) the band index $n^{*}$ is fixed by finding the $\ell=0$ eigenvalue $E_{n}^{\ell=0}$ such that the excitation distance $|(E_{n}^{0}-E_{\textrm{GS}})/N-\epsilon|$ is minimized. In this case, $\epsilon=0.2$. The scaling suggests the existence of a critical exponent $d_{c}\approx 1/2$ separating two distinct behaviors: 
\beq
\Delta E_{n^{*}}^{(0,\ell(d))}\sim N^{\omega}\,\,\textrm{with}\,\,\begin{cases}
      \omega<0 & \text{if $d<d_{c}$},\\
      \omega\approx 0 & \text{if $d=d_{c}$},\\
      \omega>0 & \text{if $d>d_{c}$}.
    \end{cases} 
\eeq
In other words, energy levels in the same band, $E_{n}^{\ell_{1}},E_{n}^{\ell_{2}}$, whose quantum number distance is $|\ell_{1}-\ell_{2}|\lesssim \sqrt{N}$ become degenerate in the limit $N\to\infty$; however, if $|\ell_{1}-\ell_{2}|\sim \sqrt{N}$, the level gap remains approximately constant with $N$, and if $|\ell_{1}-\ell_{2}|\gtrsim \sqrt{N}$ the gap actually increases with system size. A similar result applies to phase II, here shown in Fig. \ref{fig:paneldeg}(d) only for the states with even $\ell$ (because the size of phase II is smaller than that of phase I at $\xi=0.6$, the distance in this case is calculated with respect to the highest energy eigenvalue). Thus, in each band there is an infinite number of degenerate states, but there is also an infinite number of non-degenerate states. This result shows that there is an Anderson tower of states for each band $E^{\ell}_n$, $\ell=0, 1, \ldots$, up to the critical energy of the ESQPT. A similar phenomenon occurs in discrete $\mathbb{Z}_{2}$ symmetry-broken phases, where, typically, one may find pairs of degenerate levels of opposite quantum number up to a certain critical energy where an ESQPT takes place, see e.g. \cite{Khalouf2023,Cejnar2021,Perez2011,Puebla2013,Relano2014}, playing a vital role in the dynamics of many-body systems \cite{Corps2022,Corps2023arxiv,Corps2023arxivlipkin,Corps2021,Corps2023PRB}.

\subsection{Quantum quenches}\label{sec:quantumquenches}
Lastly, we focus on the dynamics of the
quantum model \eqref{eq:Hmu}. We are particularly interested in the scaling of the $\hatmath{C}_{n}$ and $\hatmath{K}_{n}$ sign operators to the $N\to\infty$ limit.

\begin{figure*}[t]
\begin{tabular}{c c}
\hspace{-1.9cm}\includegraphics[width=0.45\textwidth]{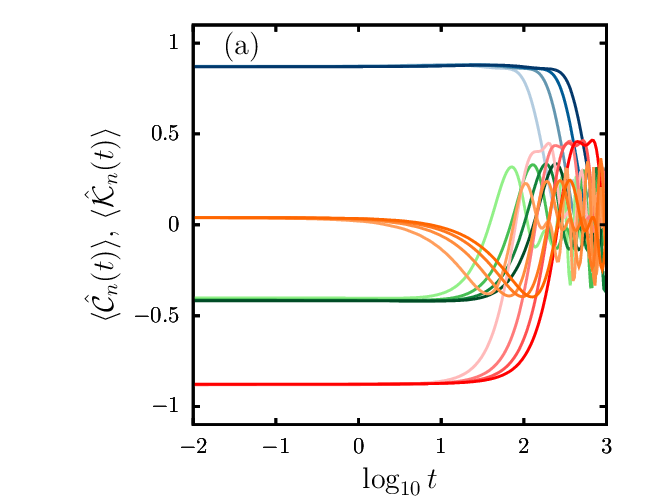} &\hspace{-0.9cm}\includegraphics[width=0.415\textwidth]{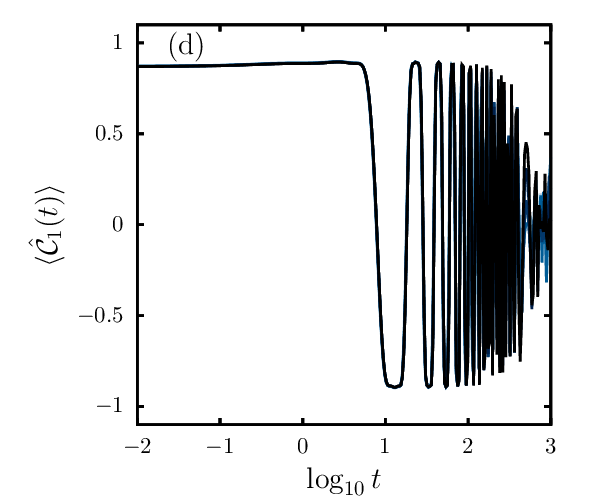}\\
\hspace{-1.2cm}\includegraphics[width=0.41\textwidth]{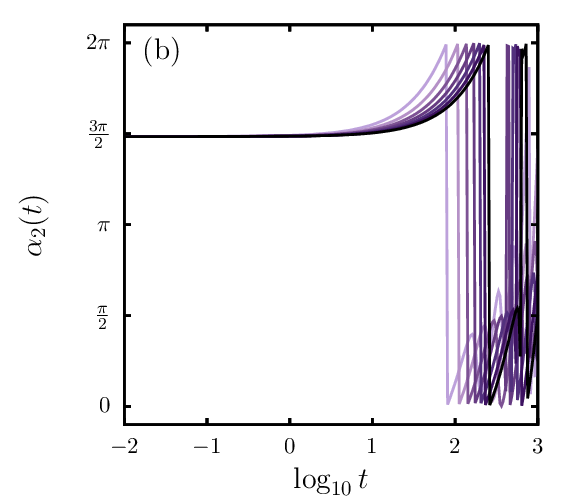} &\hspace{-0.2cm}\includegraphics[width=0.41\textwidth]{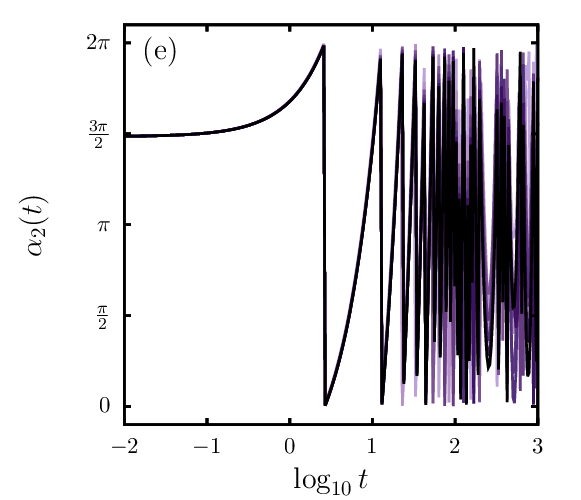}
\end{tabular}

\hspace{-1.2cm} \includegraphics[width=0.42\textwidth]{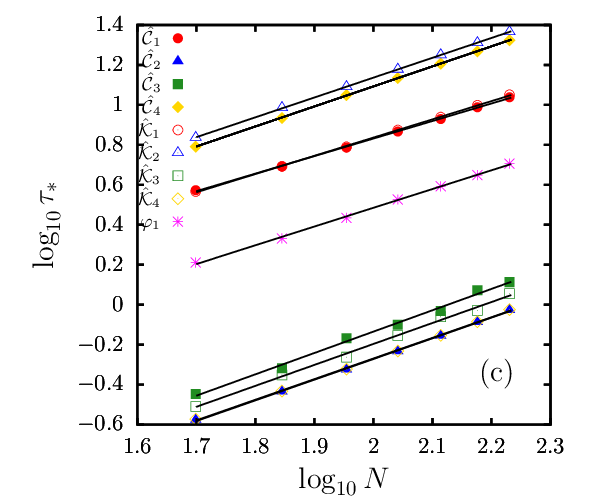}
\caption{Quench dynamics of the quantum Hamiltonian \eqref{eq:Hmu}. The Hamiltonian parameters are changed from the initial configuration $(\xi_{i}=0.6,\mu_{i}=0)$ to the final configuration $(\xi_{f}=0.8,\mu_{f})$, with $\mu_{f}=0$ [(a),(b),(c)] or $\mu_{f}=0.3$ [(d),(e)]. The initial state is \eqref{eq:coherent} with $q_{1}=-0.63$ and $q_{2}=+\sqrt{R^{2}(\xi_{i}=0.6)-q_{1}^{2}}$, with $R^{2}(\xi)$ given by \eqref{eq:R}. The coherent state is truncated to $\ell=0,1,2,3,4,5$. Quench to $\mu_{f}=0$: (a) Time evolution of $\hatmath{C}_{1}$ (blue), $\hatmath{C}_{4}$ (green), $\hatmath{K}_{1}$ (red) and $\hatmath{K}_{4}$ (orange), for $N=50,90,130,170$. Darker lines correspond to higher $N$. (b) Time evolution of the angle $\alpha_{2}$ obtained from $\langle \hat{D}_{+}^{2}(t)\rangle=r_{2}(t)e^{i\alpha_{2}t}$, for $N=50,70,\ldots,170$. (c) Scaling of the prethermalization time $\tau_{*}$ as a function of $N$ for all sign operators and the angle $\alpha_{1}$. Quench to $\mu_{f}=0.3$: (d) Time evolution of $\hatmath{C}_{1}$, for $N=70,90,\ldots,170$. (e) Time evolution of $\alpha_{2}$.  }
\label{fig:quenchesquantum}
\end{figure*}

We consider once more the $\hat{D}_{+}$ ladder operator. The sign operators $\hatmath{C}_{n}$ and $\hatmath{K}_{n}$ associated to $\hat{D}_{+}$ are calculated following Eqs. \eqref{eq:XnYn} and \eqref{eq:signs}. In order to study their quantum dynamics, we perform a quantum quench: a common and experimentally relevant procedure to force an initial state out of equilibrium. At $t=0$, an initial ground state, $\ket{\Psi_{0}(\xi_{i},\mu_{i})}$, is prepared at a set of control parameters, $(\xi_{i},\mu_{i})$. The control parameters are then suddenly changed to a final set of parameters, $(\xi_{f},\mu_{f})$, and the dynamics is generated by the final Hamiltonian, $\hat{H}_{f}\equiv \hat{H}(\xi_{f},\mu_{f})$. Following our notation for a Hamiltonian with a continuous symmetry \eqref{eq:Rsym}, the time-evolving wavefunction is given by the Schr\"{o}dinger equation,
\beq
\begin{split}
\ket{\Psi_{t}(\xi_{f},\mu_{f})}&=e^{-i\hat{H}_{f}t}\ket{\Psi_{0}(\xi_{i},\mu_{i})}\\&=\sum_{n}\sum_{\ell\in\mathbb{Z}}c_{n}^{\ell}e^{-iE_{n}(\xi_{f},\mu_{f})^{\ell}t}\ket{E_{n}^{\ell}},
\end{split}
\eeq
where $c_{n}^{\ell}\equiv \bra{E_{n}^{\ell}(\xi_{f},\mu_{f})}\ket{\Psi_{0}(\xi_{i},\mu_{i})}$ are the expansion coefficients of the initial state in the final Hamiltonian eigenbasis. We focus on quenches $(\xi_{i}=0.6,\mu_{i}=0)\to(\xi_{f}=0.8,\mu_{f})$, with $\mu_{f}=0,0.3$. The case with $\mu_{f}=0$ is chosen to illustrate the main points of our theory, while $\mu_{f}\neq 0$ provides an example of the symmetry-broken subspace of $\hat{H}_{\mu\neq 0}$ being $\mathcal{H}_{SB}=\{\emptyset\}$, which means that our theory of noncommuting constants does not apply. 

As an initial state, the coherent state \eqref{eq:coherent} is a tempting possibility, because it allows for a direct comparison of classical and quantum results. Precisely due to its classicality in the TL, such an initial state is perhaps not interesting from the quantum mechanical viewpoint. From the binomial formula $(\alpha_{+}\hat{\tau}_{+}^{\dagger}+\alpha_{-}\hat{\tau}_{-}^{\dagger})^{m}=\sum_{k=0}^{m}\binom{m}{k}(\alpha_{+}\hat{\tau}_{+}^{\dagger})^{m-k}(\alpha_{-}\hat{\tau}_{-}^{\dagger})^{k}$ it is clear that \eqref{eq:coherent} populates basis states with any possible $\ell$ quantum number, with the constraint that $N$ is conserved. A purely quantum state can be built from \eqref{eq:coherent} by truncating it to basis states with certain $\ell$. We have chosen to truncate the full coherent state to $\ell=0,1,2,3,4,5$ for this dynamical analysis. In this way, our initial state would be quantum, with quantum fluctuations, in the TL. The properties of the initial Hamiltonian enter this state through the amplitudes $\alpha_{\pm}$ in \eqref{eq:alphamasmenos}, where we have chosen $\mathbf{p}=(0,0)$, $q_{1}=-0.63$ and $q_{2}=+\sqrt{R^{2}(\xi)-q_{1}^{2}}$ according to \eqref{eq:R}. 

Our numerical results are in Fig. \ref{fig:quenchesquantum}. First, we focus on the quench with $\mu_{f}=0$. In Fig. \ref{fig:quenchesquantum}(a) we represent the quantum dynamics of $\hatmath{C}_{n}$ and $\hatmath{K}_{n}$ with $n=1$ and $4$ (similar results are found for other $n$). The quenched state only populates states of the final Hamiltonian below the ESQPT critical energy, which places it within phase I. We observe here two important results. First, at least for short times, all operators behave as a constant, with an initial value depending on the operator itself; and second, their expectation values do not coincide with what it is expected for a classical state. Thus, these results suggest that the main theoretical result proposed in Sec. \ref{sec:theory} holds in this system. 

Notwithstanding, we can also see in Fig. \ref{fig:quenchesquantum} that, as time passes by, the operators cease to be constant. According to our theory, see, e.g., \eqref{eq:H2b}, the constancy of the $\hatmath{C}_{n}$ and $\hatmath{K}_{n}$ operators can be traced back to the constancy of the angles $\alpha_{n}$ defining the dynamics of $\hat{O}^{n}=\hat{D}_{+}^{n}$; this angle is also responsible for the sign of the $\hat{X}_{n}$ and $\hat{Y}_{n}$ operators in \eqref{eq:XnYn}. In Fig. \ref{fig:quenchesquantum}(b) we test the constancy of the angle $\alpha_{2}(t)\in[0,2\pi)$ obtained from the time evolution $\langle \hat{D}_{+}^{2}(t)\rangle=r_{2}(t)e^{i\alpha_{2}(t)}$ (similar conclusions are obtained for other $n\neq 2$). Its behavior is similar to that of the $\hatmath{C}_{n}$ and $\hatmath{K}_{n}$ operators. For finite-$N$ systems, $\alpha_{2}(t)$ behaves as a constant up to a certain time, after which it starts running over the entire circumference. Our next step is to test whether this is a finite-size effect or not. We can already see that, as $N$ increases, the angle behaves as a constant during longer times.  So, we can consider that the time at which the constancy of the sign operators (as well as the symmetry-breaking angles) break down is the prethermalization time, $\tau_{*}$: before $\tau_{*}$, the time oscillating terms $\sim e^{-i(E_{n}^{\ell}-E_{n'}^{\ell'})t}$ in the Schr\"{o}dinger equation are close to vanishing, and therefore the sign operators are essentially constant. If this is a finite-size effect, and thus $\tau_* \rightarrow \infty$ when $N \rightarrow \infty$, $\tau_{*}$ has to increase with $N$. As the gap of finite-$N$ quasidegenerate eigenlevels closes as a power-law, $\Delta E_{n}^{(\ell_{1},\ell_{2})}\sim 1/N$, the prethermalization timescale is expected to increase roughly as $\tau_{*}\sim N$. This is confirmed by Fig. \ref{fig:quenchesquantum}(c), where we have represented the scaling of $\tau_{*}$ as a function of $N$. To estimate $\tau_{*}$ from the time evolution, we calculate the first value of time (up to our numerical resolution) such that $|\langle \hat{O}(t)\rangle-\langle\hat{O}(0)\rangle|>\gamma$, with a small bound $\gamma=5\times10^{-3}$, for $\hat{O}=\hatmath{C}_{n},\hatmath{K}_{n}$, and similarly for the angles. The result transparently evidences the expected power-law behavior $\tau_{*}\sim N^{b}$, with $b\approx 1$ in all cases. Thus, as the system size increases, the quantum model approaches a scenario in which $\hatmath{C}_{n}$, $\hatmath{K}_{n}$ and $\alpha_{n}$ become perfect constants, and stable symmetry-breaking equilibrium states. We have represented the scaling of the sign operators with $n=1,2,3,4$, while we only show the angle $\alpha_{2}$ as a representative case (similar results are found for the rest of $\alpha_{n}$).  

Next, we analyze the dynamics of the case $\mu_{f}=0.3$. As we have shown before, the system Hamiltonian \eqref{eq:Hmu} does not exhibit any symmetry-broken phase with $\mu_{f}\neq 0$, as there are no spectral degeneracies in the $N\to\infty$ limit. Therefore, our theory for continuous symmetry-breaking is not expected to hold in this case. In Fig. \ref{fig:quenchesquantum}(d) we represent the time evolution of the $\hatmath{C}_{1}$ operator for $\mu_{f}=0.3$. The figure clearly shows that the prethermalization time does not depend on $N$ in this case, i.e., the period of constancy does not increase as $N$ does, but remains the same with system size. The period of constancy at very short time scales should not be taken as a failure of our theory: at such short times, $\hatmath{C}_{1}$ behaves as a constant merely because the eigenlevel gaps are sufficiently small in comparison to time, but such gap does not decrease with system size and, thus, the $N\to\infty$ behavior is completely different to that of the case $\mu_{f}=0$. The angle $\alpha_{2}$ behaves in a similar way in Fig. \ref{fig:quenchesquantum}(e), which shows again that the prethermalization time of this quantity does not increase with system size. 

In short, the quantum quench dynamics is in agreement with our expectations.

\subsection{Information contained in the constants of motion}\label{sec:information}

\begin{table*}[t!]
\begin{center}
\setlength\extrarowheight{3pt}
 \begin{tabular}{||c c c c c c c c c c||} 
 \hline
 State \hspace{0.1cm} & $\langle \hatmath{C}_{1}\rangle$ \hspace{0.1cm} & $\langle \hatmath{C}_{2}\rangle$ \hspace{0.1cm} & $\langle \hatmath{C}_{3}\rangle$ \hspace{0.1cm} & $\langle \hatmath{C}_{4}\rangle$ \hspace{0.1cm} & $\langle \hatmath{K}_{1}\rangle$ \hspace{0.1cm} & $\langle \hatmath{K}_{2}\rangle$ \hspace{0.1cm} & $\langle \hatmath{K}_{3}\rangle$ \hspace{0.1cm} & $\langle \hatmath{K}_{4}\rangle$ \hspace{0.1cm} & $\langle \hat{D}_{+}/N\rangle$ \hspace{0.1cm}\\ [0.5ex] 
 \hline\hline
$\ket{\Psi}_{C}$ \hspace{0.1cm} & $1$ \hspace{0.1cm} & $-0.293$  \hspace{0.1cm} & $-1$ \hspace{0.1cm} & $-1$ \hspace{0.1cm} & $-1$ \hspace{0.1cm} & $-1$ \hspace{0.1cm} & $-1$ \hspace{0.1cm} & $0.294$ \hspace{0.1cm} & $0.680-0.713i$ \hspace{0.1cm}\\[1ex]
 \hline

$\ket{\Psi}_{Q}$ \hspace{0.1cm} & $0.333$ \hspace{0.1cm} & $-0.098$  \hspace{0.1cm} & $-1$ \hspace{0.1cm} & $0.299$ \hspace{0.1cm} & $-0.333$ \hspace{0.1cm} & $0.333$ \hspace{0.1cm} & $-1$ \hspace{0.1cm} & $0.098$ \hspace{0.1cm} & $0$ \hspace{0.1cm}\\[1ex]
 \hline

\end{tabular}
\end{center}
\caption{Expectation value of the sign operators $\hatmath{C}_{n}$, $\hatmath{K}_{n}$ ($n=1,2,3,4$) and the order parameter $\hat{D}_{+}$. The expectation value is taken with respect two different states: a classical state, given by \eqref{eq:coherent} with $q_{1}=-0.63$ at $\xi=0.6$, $\mu=0$,  and a quantum state, given by \eqref{eq:psiQ}. System size is $N=150$. Results are given up to three significant figures.}
\label{tableck}
\end{table*}

The results of the previous section show that for symmetry-breaking initial states belonging to $\mathcal{H}_{SB}$, the sign operators $\hat{\mathcal{C}}_{n}$ and $\hat{\mathcal{K}}_{n}$ behave as emergent conserved quantities that become exactly constant only in the infinite-size limit. In this section, we test if it is possible to relate the quantum fluctuations in $\alpha$ to the values $\langle \hatmath{C}_{n}\rangle$ and $\langle\hatmath{K}_{n}\rangle$, as it was proposed in Sec. \ref{sec:insight}.

In Table \ref{tableck} we report the expectation values of the sign operators up to $n=4$ for $N=150$ in the Hamiltonian with parameters $\xi=0.6$ and $\mu=0$. The expectation values are computed for two kinds of states. First, we consider a the coherent state given in \eqref{eq:coherent} with $q_{1}=-0.63$ and $(p_{1},p_{2})=(0,0)$, which we label as `classical', $\ket{\Psi}_{C}$. Second, we consider a quantum state, $\ket{\Psi}_{Q}$, built as a superposition of classical states $\ket{\Psi}_{C}$: 
\beq\label{eq:psiQ}
\ket{\Psi}_{Q}=\frac{1}{\sqrt{3}}\left[\ket{\Psi}_{C}+\hat{R}(2\pi/3)\ket{\Psi}_{C}+\hat{R}^{2}(2\pi/3)\ket{\Psi}_{C}\right],
\eeq
where $\hat{R}(\alpha)$ is the rotation in \eqref{eq:Rsym}. Observe that this state is obtained by rotating a classical state by $2\pi/3$ and $4\pi/3$ degrees. We also show the value of the order parameter, $\hat{D}_{+}$. For the classical state, we obtain $\langle \hat{D}_{+}\rangle\neq 0$, indicating that $\ket{\Psi}_{C}$ is a symmetry-breaking state. As for the sign operators up to $n=4$, many of them are already $\pm 1$ for $N=150$. However, the values of $\hatmath{C}_{2}$ and $\hatmath{K}_{4}$ seem quite distant from the expected values for a real classical state, therefore showing that the privileged direction sustained by $\ket{\Psi}_C$ has non-negligible quantum fluctuations for $N=150$. For the quantum state, we have a zero order parameter, $\langle \hat{D}_{+}\rangle=0$; yet, $\ket{\Psi}_{Q}$ is \textit{not} a symmetric state, as there are non-zero values of the sign operators. Their expectation values are clearly different from the one corresponding to $\ket{\Psi}_C$. As it consists in a superposition of three different 'classical' states with three different symmetry-breaking directions, we can expect that it will remain quantum in the thermodynamic limit, with a not well-defined value of the angle $\alpha$.

To delve into the previous facts, we take advantage of the results in Sec. \ref{sec:insight} to construct a probabilistic framework for the angle $\alpha$ in both $\ket{\Psi}_C$ and $\ket{\Psi}_Q$.
As coherent states only become classical when $N \rightarrow \infty$, we will assume that $\ket{\Psi}_C$ is a superposition of the classical states defined in Eqs. \eqref{eq:clasico1} and \eqref{eq:clasico2}, $\ket{\Psi}_C = \sum_{\alpha} c_{\alpha} \ket{\alpha}$, and therefore it gives rise to a probability density, $\rho (\alpha)$, for the angle $\alpha$. Here, we assume that distribution is Gaussian,
\beq\label{eq:rhoalpha}
\rho_{\alpha}(x)=\frac{1}{\sigma_{\alpha}\sqrt{2\pi}}e^{-(x-\overline{\alpha})^{2}/2\sigma_{\alpha}^{2}},\,\,x\in\mathbb{R},
\eeq
where $\overline{\alpha}$ is the mean of the distribution, and $\sigma_{\alpha}^{2}$ is its variance. The mean $\overline{\alpha}$ plays the role of the classical symmetry-breaking angle. Because $[\hat{q}_{k},\hat{p}_{k}]\propto 1/N$, we set $\sigma_{\alpha}^{2}=a_{\alpha}/N$, with $a_{\alpha}>0$ a free parameter. In the $N\to\infty$ limit, $\sigma_{\alpha}\to0$, and therefore $\rho_{\alpha}(x)\to\delta(x-\overline{\alpha})$, which reproduces the deterministic classical scenario. It is worth noting that the schematic states in Fig. \ref{fig:schematic}(a) are examples of these classical states. 

Because each rotated state $\hat{R}(\alpha)\ket{\Psi}_{C}$ is a classical state, it is associated to a Gaussian distribution $\rho_{\alpha}$ of the type \eqref{eq:rhoalpha}. Therefore, the angle distribution for the quantum state \eqref{eq:psiQ} reads
\beq\label{eq:rhosuperposition}
\rho(x)=\frac{1}{3}[\rho_{\alpha}(x)+\rho_{\alpha+2\pi/3}(x)+\rho_{\alpha+4\pi/3}(x)].
\eeq
Each distribution entering \eqref{eq:rhosuperposition} has its own mean, $\overline{\alpha},\overline{\alpha}+2\pi/3,\overline{\alpha}+4\pi/3$, and its own variance, $\sigma^{2}_{\alpha'}=a_{\alpha'}/N$.  We note that quantum superpositions of classical states are not classical regardless of $N$, as the variance of \eqref{eq:rhosuperposition} does not vanish even in the infinite-size limit, when the distribution reduces to $\rho(x)=[\delta(x-\overline{\alpha})+\delta(x-\overline{\alpha}-2\pi/3)+\delta(x-\overline{\alpha}-4\pi/3)]/3$. The schematic superposition in Fig. \ref{fig:schematic}(b) is an example of these states. 

Under these circumstances, the expectation values of the $\hatmath{C}_{n}$ and $\hatmath{K}_{n}$ operators follow from \eqref{eq:quantum2} with the prescription that $\alpha$ is a continuous random variable, i.e,
\begin{subequations}
\label{eq:integralsCK}
\begin{align}
&\bra{\Psi} \hat{\mathcal{C}}_n \ket{\Psi} = \int_{-\infty}^{\infty}\textrm{d}\alpha\,\rho(\alpha) \textrm{sign}[\cos(\alpha n)], \\
&\bra{\Psi} \hat{\mathcal{K}}_n \ket{\Psi} = \int_{-\infty}^{\infty}\textrm{d}\alpha\,\rho(\alpha) \textrm{sign}[\sin(\alpha n)],
\end{align}
\end{subequations}
where $\ket{\Psi}$ may be a classical or a quantum state. For $\ket{\Psi}_{C}$, in the $N\to\infty$ limit, one trivially obtains $\langle \hatmath{C}_{n}\rangle_{C}=\textrm{sign}[\cos(n\overline{\alpha})]$, $\langle \hatmath{K}_{n}\rangle_{C}=\textrm{sign}[\sin(n\overline{\alpha})]$. Yet, for finite-$N$ systems the non-zero $\sigma_{\alpha}$ means that deviations from these values may be relevant. For $\ket{\Psi}_{Q}$, the equivalent $N\to\infty$ results are $\langle\hat{\mathcal{C}}_{n}\rangle_{Q}=\{\textrm{sign}[\cos(n\overline{\alpha})]+\textrm{sign}[\cos(n(\overline{\alpha}+2\pi/3))]+\textrm{sign}[\cos(n(\overline{\alpha}+4\pi/3))]\}/3$ and $\langle\hat{\mathcal{K}}_{n}\rangle_{Q}=\{\textrm{sign}[\sin(n\overline{\alpha})]+\textrm{sign}[\sin(n(\overline{\alpha}+2\pi/3))]+\textrm{sign}[\sin(n(\overline{\alpha}+4\pi/3))]\}/3$, with deviations from these values appearing for $N<\infty$ systems. In any case, it is worth noting that for a classical state, the only possible values of $\hatmath{C}_{n}$ and $\hatmath{K}_{n}$, in the $N\to\infty$ limit, are $-1,0,1$, while for the quantum state \eqref{eq:psiQ} the possible values are $-1,-1/3,1/3,1$. The parameter $a_{\alpha}$ in $\sigma_{\alpha}$ is obtained by solving the equation $\bra{\Psi}\hatmath{C}_{2}\ket{\Psi}=-0.293$ for $\ket{\Psi}_{C}$ and $N=150$ (see Table \ref{tableck}), which yields a value $a_{\alpha}=0.473472$. 

\begin{figure}[h!]
\begin{tabular}{c}
\hspace{-0.7cm}\includegraphics[width=0.48\textwidth]{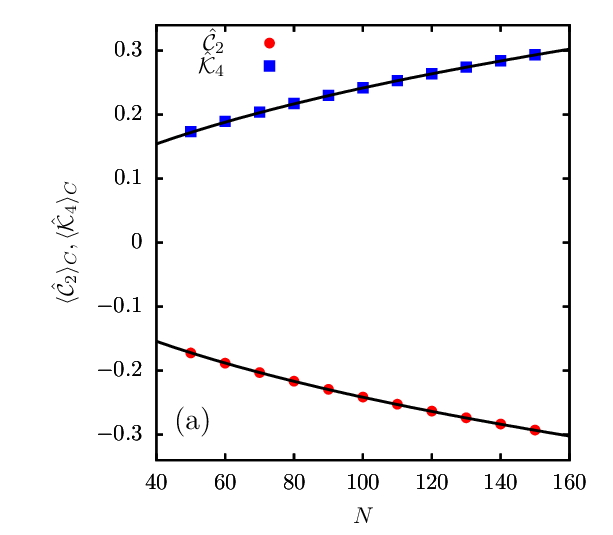}\\
\hspace{-1.2cm}\includegraphics[width=0.51\textwidth]{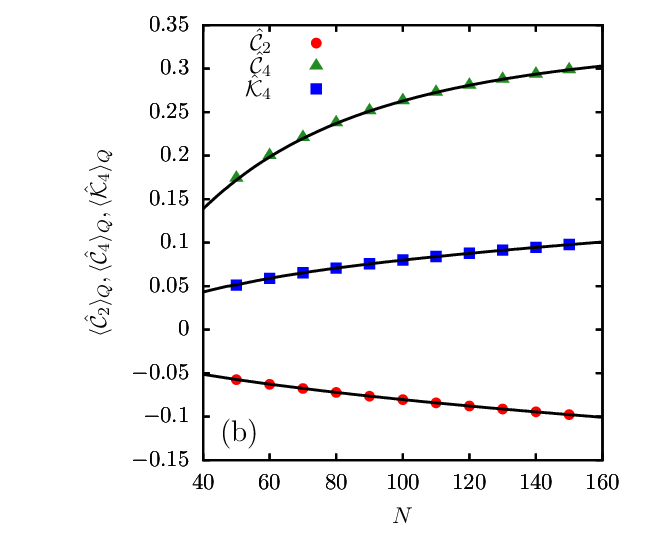}
\end{tabular}
\caption{Scaling of the expectation value of the sign operators with system size. (a) Scaling of $\hatmath{C}_{2}$ and $\hatmath{K}_{4}$ for a classical state, $\ket{\Psi}_{C}$, of the type \eqref{eq:coherent} with $q_{1}=-0.63$, $q_{2}=\sqrt{R^{2}(\xi)-q_{1}^{2}}$, with $R^{2}(\xi)$ given in \eqref{eq:R}, $(p_{1},p_{2})=(0,0)$ and $(\xi=0.6,\mu=0)$. (b) Scaling of $\hatmath{C}_{2}$, $\hatmath{C}_{4}$ and $\hatmath{K}_{4}$ for a quantum superposition \eqref{eq:psiQ} with $\ket{\Psi}_{C}$ as in (a). Black lines represent the theoretical predictions of \eqref{eq:integralsCK}.}
\label{fig:scalingCnKn}
\end{figure}

To test this model, we represent in Fig. \ref{fig:scalingCnKn} the scaling of the questionable sign operators for the classical state [(a)] and for the quantum state [(b)]. The theoretical predictions of \eqref{eq:integralsCK} are represented by black lines. The theoretical result matches the numerical data exceptionally well, taking into account the simplicity of our framework. In the $N\to\infty$ limit, \eqref{eq:integralsCK} yields $\langle\hatmath{C}_{2}\rangle_{C}=-1$ and $\langle\hatmath{K}_{4}\rangle_{C}=1$ for the classical state, and $\langle\hatmath{C}_{2}\rangle_{Q}=-1/3$, $\langle\hatmath{C}_{4}\rangle_{Q}=1/3$, and $\langle\hatmath{K}_{4}\rangle_{C}=1/3$ for the quantum state. Clearly, the rate of convergence to the $N\to\infty$ result depends on each operator. 

These results show that the assumption that the angle distribution of a quantum state can be roughly described by a classical probability distribution works very well. Together with the results of the previous section, we have learned that in the $N\to\infty$ limit there is information about the initial condition that is conserved by the dynamics of continuous symmetry-breaking Hamiltonians. Thus, the dynamical spreading of this information gets completely frozen in the thermodynamic limit. As the main conclusion, we highlight that the knowledge of $\langle \hatmath{C}_{n}\rangle$ and $\langle \hatmath{K}_{n}\rangle$ allows us to infer the form of the angle distribution $\rho_{\alpha}$ and the separation of classical directions.

\section{Conclusions}\label{sec:conclusions}
In this paper we have presented a theory for the dynamics of symmetry-breaking phases in systems with a continuous (rotational) symmetry. Our main result is that in the symmetry-broken phase one can define an extensive number of quantum operators that act as constants of motion. Such operators can be constructed directly from the complex order parameter of the phase transition, and equilibrium states depend on its expectation (constant) values in the symmetry-broken phase. In addition, these operators are related to the symmetry-breaking angle of the polar representation of the order parameter, to which one may attribute classical significance. Our numerical results are illustrated with a fully-connected system, the two-dimensional limit of the vibron model. Importantly, quantum quenches show that as the system size increases so does the prethermalization time, which defines the timescale below which the proposed operators are constant. In our case, this timescale grows as a power-law with $N$, which is in turn connected to the closure of energy gaps in the symmetry-breaking phase; only in the infinite-size limit do these energy distances vanish, and so the operators $\hatmath{C}_{n}$ and $\hatmath{K}_{n}$ become perfect constants in this limit. Additionally, we propose a statistical model describing the the expectation values of these observables relying on a probability distribution over the classical symmetry-breaking angle of the order parameter. Our results imply that this angle is a quantum magnitude, suffering from quantum fluctuations, under realistic circumstances.

\begin{acknowledgments}
The authors are thankful to J. M. Arias and F. P\'{e}rez-Bernal for useful discussions. This work has been supported by the Spanish grants PGC-2018-094180-B-I00,  PID2019-106820RB-C21 and PID2022-136285NB-C31, funded by Ministerio de Ciencia e Innovaci\'{o}n/Agencia Estatal de Investigaci\'{o}n MCIN/AEI/10.13039/501100011033 and FEDER "A Way of Making Europe". A. L. C. acknowledges financial support from `la Caixa' Foundation (ID 100010434) through the fellowship LCF/BQ/DR21/11880024.
 \end{acknowledgments}

\appendix
\section{Some proofs}\label{sec:appendixI}
We want to show that if the order parameter $\hat{O}$ satisfies Eq. \eqref{eq:propertyO}, and $E_{m}^{2i}=E_{m}^{2j}$ for all $i,j\in\mathbb{Z}$, then the operators $\hat{\mathcal{C}}_{2n-1}$ and $\hat{\mathcal{K}}_{2n-1}$ are not constants of motion. For this, it is sufficient to show that the probability conservation in Eq. \eqref{eq:propertyprobability} is not satisfied. Consider an initial state $\ket{\psi}=\sum_{n,\ell}c_{n}^{\ell}\ket{E_{n}^{\ell}}$. The instantaneous expectation value of a general observable, not necessarily of the form Eq. \eqref{eq:propertyO}, is
\beq
\begin{split}
&\langle\hat{O}(t)\rangle=\sum_{n}\widetilde{\sum_{k,\ell}}(c_{n}^{k})^{*}c_{n}^{\ell}\bra{E_{n}^{k}}\hat{O}\ket{E_{n}^{\ell}}\\
&+\sum_{n}\overline{\sum_{k,\ell}}(c_{n}^{k})^{*}c_{n}^{\ell}e^{-i(E_{n}^{\ell}-E_{n}^{k})t}\bra{E_{n}^{k}}\hat{O}\ket{E_{n}^{\ell}}\\
&+\sum_{n\neq m}\sum_{k,\ell}(c_{m}^{k})^{*}c_{n}^{\ell}e^{-i(E_{n}^{\ell}-E_{m}^{k})t}\bra{E_{m}^{k}}\hat{O}\ket{E_{n}^{\ell}}.
\end{split}
\eeq
Here, the $\widetilde{\sum_{k,\ell}}$ summation is performed over pairs of degenerate eigenlevels (hence the vanishing undulatory term), while the $\overline{\sum_{k,\ell}}$ summation concerns pairs of non degenerate eigenlevels. Let us assume that $E_{n}^{\ell+2j}=E_{n}^{\ell}$, $j\in\mathbb{Z}$. If the operator $\hat{O}=\hat{X}_{2n-1}$ or $\hat{O}=\hat{Y}_{2n-1}$ with $n\in\mathbb{N}$, the first summation vanishes because states with $\ell$ and $k=\ell+2j$ are not connected by $\hat{O}$. The second and third summations are not identically zero, but its long-time average vanishes. Therefore, for this choice of $\hat{O}$, 
\beq
\overline{\langle \hat{O}\rangle}=\lim_{\tau\to\infty}\frac{1}{\tau}\int_{0}^{\tau}\textrm{d}t\,\langle\hat{O}(t)\rangle=0.
\eeq
This means that $\langle \hat{O}(t)\rangle$ fluctuates around zero, and therefore its sign changes over time. Thus, hypothesis (H3) is not verified, and thus $\hat{\mathcal{C}}_{2n-1}$ and $\hat{\mathcal{K}}_{2n-1}$ are not constants of motion.

\section{Holstein-Primakoff transformation}\label{sec:appendixII}
The $U(3)$ vibron model is constructed as a linear
combination of the Casimir invariants of the two subalgebras chains, namely
operators $U(2)$ and $SO(3)$ \cite{PerezBernal2008}. From the $9$ one-body
elements of $U(3)$, the $\hat{\sigma} $ boson can be eliminated by means of the
Holstein-Primakoff (HP) transformation:
\begin{equation}
\hat{\tau} _{\mu }^{\dagger }\hat{\tau} _{\nu }=\hat{a}_{\mu }^{\dagger }\hat{a}_{\nu },\label{uno}
\end{equation}
\begin{equation}
\hat{\tau} _{\mu }^{\dagger }\hat{\sigma} =\sqrt{N}\hat{a}_{\mu }^{\dagger }\sqrt{1-\frac{
\sum_{\nu=+,- }\hat{a}_{\nu }^{\dagger }\hat{a}_{\nu }}{N}},  \label{dos}
\end{equation}
\begin{equation}
\hat{\sigma} ^{\dagger }\hat{\sigma} =N-\sum_{\nu=+,- }\hat{a}_{\nu }^{\dagger }\hat{a}_{\nu }
\label{tres}
\end{equation}
where $\mu=\in\{+,-\}$. Observe that the circular bosons $\hat{\tau}_{\pm }$ are
mapped to bosons $\hat{a}_{\pm }$, and the cost of the $\hat{\sigma} $-boson
elimination is an infinite boson expansion in \eqref{dos}. The presence of the 
$1/N$ term inside the square-root has been employed to truncate
the Taylor expansion to the first order terms. However, the expectation
value $\hat{n}_{a}=\sum_{\nu }\left\langle \hat{a}_{\nu }^{\dagger }\hat{a}_{\nu
}\right\rangle/N$, which has been used as an order parameter to
characterize the second order phase transtion of the vibron model \cite%
{PerezBernal2008}, acquires a finite value in the symmetry-broken phase
precluding an effective truncation. In order to proceed in the analysis of
the infinite-size limit, we propose a shift transformation
\begin{equation}
\hat{b}_{\pm }=\sqrt{N}\alpha _{\pm }-\hat{a}_{\pm },
\end{equation}
where $\alpha _{\pm }\in \mathbb{C}$ and $\hat{b}_{\pm }$ is a new set of boson
annihilation operators whose vacuum is the coherent state
\begin{equation}
\left\vert \Phi \right\rangle _{\textrm{HP}}=\frac{1}{\mathcal{N}}e^{\sqrt{N}\left(
\alpha _{+}\hat{a}_{+}^{\dagger }+\alpha _{-}\hat{a}_{-}^{\dagger }\right) }\left\vert
0\right\rangle,  \label{eq:coherentHP}
\end{equation}
where $\mathcal{N}^{2}=e^{N\left( \left\vert \alpha _{+}\right\vert
^{2}+\left\vert \alpha _{-}\right\vert ^{2}\right) }$ is such that $\bra{\Phi}\ket{\Phi}=1$. 

Expectation values of any polynomial of the operators $\{ \hat{a}_{\pm
},\hat{a}_{\pm }^{\dagger }\} $ amount to the replacements $\hat{a}_{\pm
}\rightarrow \sqrt{N}\alpha _{\pm }$, and $\hat{a}_{\pm }^{\dagger }\rightarrow \sqrt{N}\alpha _{\pm }^{*}$.

The coherent state \eqref{eq:coherentHP} can be expanded in the HP orthonormal basis,
\begin{equation}
\left\vert n_{+},n_{-}\right\rangle _{\textrm{HP}}=\frac{1}{\sqrt{n_{+}!n_{-}!}}%
\hat{a}_{+}^{\dagger n_{+}}\hat{a}_{-}^{\dagger n_{-}}\left\vert 0\right\rangle,
\label{eq:HPorthonormalbasis}
\end{equation}
as
\begin{equation}
\left\vert \Phi \right\rangle _{\textrm{HP}}=\sum_{m=0}^{N}\frac{1}{m!}\sqrt{N^{m}}\left( \alpha
_{+}\hat{a}_{+}^{\dagger }+\alpha _{-}\hat{a}_{-}^{\dagger }\right) ^{m}\left\vert
0\right\rangle.
\end{equation}
Making the relation between the HP basis \eqref{eq:HPorthonormalbasis} and the original
orthonormal basis,
\beq
\ket{n_{+}\,n_{-}}=\frac{\hat{\sigma}^{\dagger N-n_{+}-n_{-}}\hat{\tau_{+}}^{\dagger n_{+}}\hat{\tau}_{-}^{\dagger n_{-}}}{\sqrt{(N-n_{+}-n_{-})!n_{+}!n_{-}!}}\ket{0},
\eeq
the HP coherent state \eqref{eq:coherentHP} state takes the form of Eq. \eqref{eq:coherent} in the main text.

The coherent state \eqref{eq:coherentHP} makes it possible to derive the following energy functional from the quantum Hamiltonian:
\beq\label{eq:classicalHP}\begin{split}
H_{\mu}&=\frac{\bra{\Phi}\hat{H}_{\mu}\ket{\Phi}}{N}=(1-\xi)(|\alpha_{+}|^{2}+|\alpha_{-}|^{2})+\xi\\&
-2\xi(1-|\alpha_{+}|^{2}-|\alpha_{-}|^{2})(|\alpha_{+}|^{2}+|\alpha_{-}|^{2}-\alpha_{+}\alpha_{-}-\alpha_{+}^{*}\alpha_{-}^{*})\\&
-\xi(|\alpha_{+}|^{4}+|\alpha_{-}|^{4}-2|\alpha_{+}|^{2}|\alpha_{-}|^{2})\\&
+\mu(|\alpha_{+}|^{2}-|\alpha_{-}|^{2}),
\end{split}
\eeq
with the constraint
\beq\label{eq:constraint}
\frac{\bra{\Phi}\hat{n}_{a}\ket{\Phi}_{\textrm{HP}}}{N}=|\alpha_{+}|^{2}+|\alpha_{-}|^{2}\leq 1.
\eeq
Since we are interested in the $N\to\infty$ limit, multiplicative factors $N/(N-1)$ in \eqref{eq:classicalHP} have been set to unity for simplicity.

In order to arrive to the classical Hamiltonian in \eqref{eq:classicalham}, we need to express the coherent state coefficients in terms of the canonical variables $(q_{1},p_{1},q_{2},p_{2})$. Taking into account Eqs. \eqref{eq:taumasmenos} and \eqref{eq:tau12}, we find
\begin{subequations}\begin{align}
&\hat{q}_{k}=\frac{1}{\sqrt{2N}}(\hat{\tau}_{k}^{\dagger}+\hat{\tau}_{k}),\\&
\hat{p}_{k}=\frac{i}{\sqrt{2N}}(\hat{\tau}_{k}^{\dagger}-\hat{\tau}_{k}),
\end{align}\end{subequations}
with $k=1,2$, and
\begin{subequations}\begin{align}
    &\hat{\tau}_{1}=\frac{1}{\sqrt{2}}(\hat{\tau}_{-}-\hat{\tau}_{+}),\\&
    \hat{\tau}_{2}=-\frac{i}{\sqrt{2}}(\hat{\tau}_{+}+\hat{\tau}_{-}).
\end{align}\end{subequations}
Expressing the $\hat{q}_{k}$ and $\hat{p}_{k}$ operators in terms of the circular bosons $\hat{\tau}_{\pm}$, and taking the expectation values
\beq
\begin{split}
q_{k}\equiv \bra{\Phi}\hat{q}_{k}\ket{\Phi}_{\textrm{HP}},\,\,
p_{k}\equiv \bra{\Phi}\hat{p}_{k}\ket{\Phi}_{\textrm{HP}},
\end{split}\eeq
we derive the following system of equations:
\beq\label{eq:systemHP}
\begin{cases}
    q_{1}=\Re(\alpha_{-}-\alpha_{+}),\\
    q_{2}=\Im(\alpha_{+}+\alpha_{-}),\\
    p_{1}=\Im(\alpha_{-}-\alpha_{+}),\\
    p_{2}=-\Re(\alpha_{+}+\alpha_{-}).
\end{cases}
\eeq
Solving for $\alpha_{\pm}=\alpha_{\pm}(\mathbf{q},\mathbf{p})$ gives the solution in \eqref{eq:alphamasmenos}. Finally, substituting the solution \eqref{eq:alphamasmenos} in \eqref{eq:classicalHP} yields the classical Hamiltonian \eqref{eq:classicalham}. Also, note that the condition in \eqref{eq:constraint} constrains the classical dynamics to the subset of $\mathbb{R}^{4}$ defined by $\Omega^{2}=q_{1}^{2}+p_{1}^{2}+q_{2}^{2}+p_{2}^{2}\leq 2$.

The classical limit of the operator $\hat{D}_{+}$, which we use in the main text, can be calculated in a similar fashion, namely:
\beq\begin{split}
\frac{\bra{\Phi}\hat{D}_{+}\ket{\Phi}_{\textrm{HP}}}{N}&=\sqrt{2(1-|\alpha_{+}|^{2}-|\alpha_{-}|^{2})}(\alpha_{+}^{*}-\alpha_{-})\\&=-\sqrt{2-\Omega^{2}}(q_{1}+iq_{2}).
\end{split}
\eeq


\begin{thebibliography}{100}

\bibitem{Arodz2012} H. Arodz, J. Dziarmaga, and W. H. Zurek, \textit{Patterns of Symmetry Breaking} (Springer Netherlands, Dordrecht, 2012) vol 127.

\bibitem{Beekman2019} A. J. Beekman, L. Rademaker, and J. van Wezel\textit{An introduction to spontaneous symmetry breaking}, SciPost Phys. Lect. Notes \textbf{11} (2019).

\bibitem{Landau1937} L. D. Landau, \textit{On the theory of phase transitions}, I. Z. Eksp. Teor. Fiz. \textbf{11}, 19 (1937).

\bibitem{Landau1950} L. D. Landau and V. L. Ginzburg, \textit{On the theory of superconductivity}, Zh. Eksp. Theor. Fiz. \textbf{20}, 1064 (1950).

\bibitem{Englert1964} F. Englert and R. Brout, \textit{Broken Symmetry and the Mass of Gauge Vector Mesons}, Phys. Rev. Lett. \textbf{13}, 321 (1964).

\bibitem{Higgs1964} P. W. Higgs \textit{Broken Symmetries and the Masses of Gauge Bosons}, Phys. Rev. Lett. \textbf{13}, 508 (1964).

\bibitem{Guralnik1964} G. S. Guralnik, C. R. Hagen, and T. W. B. Kibble, \textit{Global Conservation Laws and Massless Particles}, Phys. Rev. Lett. \textbf{13}, 585 (1964).

\bibitem{Albrecht1982} A. Albrecht and P. J. Steinhardt, \textit{Cosmology for Grand Unified Theories with Radiatively Induced Symmetry Breaking}, Phys. Rev. Lett. \textbf{48}, 1220 (1982).

\bibitem{Kibble1976} T. W. B. Kibble, \textit{Topology of cosmic domains and strings}, J. Phys. A. \textbf{9}, 1387 (1976).

\bibitem{Nambu1960} Y. Nambu, \textit{Quasi-Particles and Gauge Invariance in the Theory of Superconductivity}, Phys. Rev. \textbf{117}, 648 (1960).

\bibitem{Goldstone1961} J. Goldstone, \textit{Field theories with 'Superconductor' solutions}, Il Nuovo Cimento, \textbf{19}, 154 (1961).

\bibitem{Yang1962} C. N. Yang, \textit{Concept of off-diagonal long-range order and the quantum phases of liquid He and of superconductors}, Rev. Mod. Phys. \textbf{34}, 694 (1962).

\bibitem{Anderson1952} P. W. Anderson, \textit{An Approximate Quantum Theory of the Antiferromagnetic Ground State}, Phys. Rev. \textbf{86}, 694 (1952).

\bibitem{Bernu1992} B. Bernu, C. Lhuillier, and L. Pierre, \textit{Signature of N\'eel Order in Exact Spectra of Quantum Antiferromagnets on Finite Lattices}, Phys. Rev. Lett. \textbf{69}, 2590 (1992).

\bibitem{Azaria1993} P. Azaria, B. Delamotte, and D. Mouhanna, \textit{Spontaneous Symmetry Breaking in Quantum Frustrated Antiferromagnets}, Phys. Rev. Lett. \textbf{70}, 2483 (1993).

\bibitem{Bernu1994} B. Bernu, P. Lecheminant, C. Lhuillier, and L. Pierre, \textit{Exact spectra, spin susceptibilities, and order parameter of the quantum Heisenberg antiferromagnet on a triangular lattice}, Phys. Rev. B \textbf{50}, 10048 (1994).

\bibitem{Koma1994} T. Koma and H. Tasaki, \textit{Symmetry Breaking and Finite Size Effects in Quantum Many-Body Systems}, J. Stat. Phys. \textbf{76}, 745 (1994).

\bibitem{Tasaki2019} H. Tasaki, \textit{Long-Range Order, "Tower" of States, and Symmetry Breaking in Lattice Quantum Systems}, J. Stat. Phys. \textbf{174}, 735 (2019).

\bibitem{Gersch1963} H. A. Gersch and G. C. Knollman, \textit{Quantum Cell Model for Bosons}, Phys. Rev. \textbf{129}, 959 (1963).

\bibitem{Sachdev} S. Sachdev, \textit{Quantum Phase Transitions}, (Cambridge University Press, 2011).

\bibitem{Fisher1989} M. P. A. Fisher, P. B. Weichman, G. Grinstein, and D. S. Fisher, \textit{Boson localization and the superfluid-insulator transition}, Phys. Rev. B \textbf{40}, 546 (1989).

\bibitem{Greiner2002} M. Greiner, O. Mandel, T. Esslinger, T. W. Ha\"ansch, and I. Bloch, \textit{Quantum phase transition from a superfluid to a Mott insulator in a gas of ultracold atoms}, Nature \textbf{415}, 39 (2002).

\bibitem{Dyson1978} F. J. Dyson, E. H. Lieb, and B. Simon, \textit{Phase transitions in quantum spin systems with isotropic and non-isotropic interactions}, J. Stat. Phys. \textbf{18}, 335 (1978).

\bibitem{Mazurenko2017} A. Mazurenko, C. S. Chiu, G. Ji, M. F. Parsons, M. Kan\'asz-Nagy, R. Schmidt, F. Grusdt, E. Demler, D. Greif, and M. Greiner, \textit{A cold-atom Fermi-Hubbard antiferromagnet}, Nature \textbf{545}, 462 (2017).

\bibitem{Larson2017} J. Larson and E. K. Irish, \textit{Some remarks on 'superradiant' phase transitions in light-matter systems}, J. Phys. A: Math. Theor. \textbf{50}, 174002 (2017).

\bibitem{Corps2021} A. L. Corps and A. Rela\~{n}o, \textit{Constant of Motion Identifying Excited-State Quantum Phases}, Phys. Rev. Lett. \textbf{127}, 130602 (2021)

\bibitem{Corps2022} A. L. Corps and A. Rela\~{n}o, \textit{Dynamical and excited-state quantum phase transitions in collective systems}, Phys. Rev. B \textbf{106}, 024311 (2022).

\bibitem{Corps2023} A. L. Corps and A. Rela\~{n}o, \textit{Theory of Dynamical Phase Transitions in Quantum Systems with Symmetry-Breaking Eigenstates}, Phys. Rev. Lett. \textbf{130}, 100402 (2023).

\bibitem{Corps2023arxiv} A. L. Corps and A. Rela\~{n}o, \textit{General theory for discrete symmetry-breaking equilibrium states}, arXiv:2303.18020.

\bibitem{Khalouf2021} J. Khalouf-Rivera, F. Pérez-Bernal, and M. Carvajal, \textit{Excited state quantum phase transitions in the bending spectra of
molecules}, J. Quant. Spectrosc. Radiat. Transfer \textbf{261}, 107436
(2021).

\bibitem{PerezBernal2005} F. Pérez-Bernal, L. F. Santos, P. H. Vaccaro, and F. Iachello,
\textit{Spectroscopic signatures of nonrigidity: Algebraic analyses of
infrared and Raman transitions in nonrigid species}, Chem. Phys.
Lett. \textbf{414}, 398 (2005).

\bibitem{Iachello2003} F. Iachello, F. Pérez-Bernal, and P. H. Vaccaro, \textit{A novel algebraic scheme for describing nonrigid molecules}, Chem. Phys.
Lett. \textbf{375}, 309 (2003).

\bibitem{Iachello1996} F. Iachello and S. Oss, \textit{Algebraic approach to molecular spectra:
Two dimensional problems}, J. Chem. Phys. \textbf{104}, 6956 (1996).

\bibitem{PerezBernal2008} F. P\'{e}rez-Bernal and F. Iachello, \textit{Algebraic approach to two-dimensional systems: Shape phase transitions, monodromy, and thermodynamic quantities},  Phys. Rev. A \textbf{77}, 032115 (2008).

\bibitem{Khalouf2022} J. Khalouf-Rivera, F. P\'{e}rez-Bernal and M. Carvajal, \textit{Anharmonicity-induced excited-state quantum phase transition in the symmetric phase of the two-dimensional limit of the vibron model}, Phys. Rev. A \textbf{105}, 032215 (2022).

\bibitem{PerezBernal2010} F. P\'{e}rez-Bernal and O. \'{A}lvarez-Bajo, \textit{Anharmonicity effects in the bosonic U(2)-SO(3) excited-state quantum phase transition}, Phys. Rev. A \textbf{81}, 050101(R) (2010).

\bibitem{Novotny2023} J. Novotn\'{y} and P. Str\'{a}nsk\'{y}, \textit{Relative asymptotic oscillations of the out-of-time-ordered correlator as a quantum chaos indicator},
Phys. Rev. E \textbf{107}, 054220 (2023).

\bibitem{Jaynes1957a} E. T. Jaynes, \textit{Information theory and statistical mechanics}, Phys. Rev. {\bf 106}, 620 (1957).

\bibitem{Jaynes1957b} E. T. Jaynes, \textit{Information theory and statistical mechanics. II}, Phys. Rev. {\bf 108}, 171 (1957).

\bibitem{Rigol2007} M. Rigol, V. Dunjko, V. Yurovsky, and M. Olshanii, \textit{Relaxation in a completely integrable many-body quantum system: An ab initio study of the dynamics of the highly excited states of 1D lattice
hard-core bosons}, Phys. Rev. Lett. {\bf 98}, 050405 (2007).

\bibitem{Tinkhambook} M. Tinkham, Introduction to Superconductivity, 2nd Edition, Dover Publications Inc (2004).

\bibitem{Caprio2008} M. A. Caprio, P. Cejnar and F. Iachello. Excited state quantum phase transitions in many-body systems. \textit{Ann. Phys. (N.Y.)} \textbf{323}, 1106 (2008).

\bibitem{Cejnar2021} P. Cejnar, P. Str\'{a}nsk\'{y}, M. Macek and M. Kloc. Excited-state quantum phase transitions. \textit{J. Phys. A: Math. Theor.} \textbf{54}, 133001 (2021).

\bibitem{Holstein1940} T. Holstein and H. Primakoff, \textit{Field Dependence of the Intrinsic Domain Magnetization of a Ferromagnet}, Phys. Rev. \textbf{58}, 1098 (1940).

\bibitem{Khalouf2023} J. Khalouf-Rivera, Q. Wang, L. F. Santos, J. E. Garc\'{i}a-Ramos, M. Carvajal, and F. P\'{e}rez-Bernal, \textit{Degeneracy in excited state quantum phase transitions of two-level bosonic models and its influence on system dynamics}, arXiv:2303.16551.

\bibitem{Perez2011} P. P\'erez-Fern\'andez, P. Cejnar, J. M. Arias, J. Dukelsky, J. E. Garc\'{\i}a-Ramos, and A. Rela\~no, {\em Quantum quench influenced by an excited-state phase transition}, Phys. Rev. A {\bf 83}, 033802 (2011).

\bibitem{Puebla2013} R. Puebla, A. Rela\~{n}o, and J. Retamosa, \textit{Excited-state phase transition leading to symmetry-breaking steady states in the Dicke model}, Phys. Rev. A \textbf{87}, 023819 (2013).

\bibitem{Relano2014} A. Rela\~{n}o, J. Dukelsky, P. P\'{e}rez-Fern\'{a}ndez, and J. M. Arias, \textit{Quantum phase transitions of atom-molecule Bose mixtures in a double-well potential}, Phys. Rev. E \textbf{90}, 042139 (2014).

\bibitem{Corps2023arxivlipkin} A. L. Corps, P. P\'{e}rez-Fern\'{a}ndez, and A. Rela\~{n}o, \textit{Relaxation time as a control parameter for exploring dynamical phase diagrams}, Phys. Rev. B \textbf{108}, 174305 (2023).

\bibitem{Corps2023PRB} A. L. Corps, P. Str\'{a}nsk\'{y}, and P. Cejnar, \textit{Mechanism of dynamical phase transitions: The complex-time survival amplitude}, Phys. Rev. B \textbf{107}, 094307 (2023).


\end{thebibliography}
\end{document}